\documentclass[submission, Phys]{SciPost}
\hypersetup{
    colorlinks=true,
    linkcolor=black,
    citecolor=blue,
    urlcolor=blue
}
\usepackage[utf8]{inputenc}
\usepackage[dvipsnames]{xcolor}
\usepackage{amsthm, amssymb}
\usepackage{amsmath}
\usepackage{braket}
\usepackage{txfonts,stmaryrd}
\usepackage{graphicx}
\usepackage{enumerate}

\usepackage{bm}
\usepackage{bbm}
\usepackage{booktabs} 
\usepackage{mdframed}
\usepackage[normalem]{ulem}

\begin{document}

\begin{center}
{\Large \textbf{Complexity transitions in chaotic quantum systems: Nonstabilizerness, entanglement, and fractal dimension in SYK and random matrix models}}
\end{center}

\vspace{2mm}

\begin{center}
\textbf{Gopal Chandra Santra\textsuperscript{1,2,3*}, 
Alex Windey\textsuperscript{1,2\dag}, 
Soumik Bandyopadhyay\textsuperscript{1,2\ddag}, 
Andrea Legramandi\textsuperscript{1,2\S}, 
Philipp Hauke\textsuperscript{1,2\#}}
\end{center}

\vspace{1mm}

\begin{center}
\textsuperscript{1} Pitaevskii BEC Center, CNR-INO and Department of Physics, University of Trento, Via Sommarive 14, I-38123 Trento, Italy\\
\textsuperscript{2} INFN-TIFPA, Trento Institute for Fundamental Physics and Applications, Via Sommarive 14, I-38123 Trento, Italy\\
\textsuperscript{3} Kirchhoff-Institut f\"ur Physik, Universit\"at Heidelberg, Im Neuenheimer Feld 227, 69120 Heidelberg, Germany\\[1mm]
*\textcolor{blue}{gopalchandra.santra@unitn.it}, \dag \textcolor{blue}{alex.windey@unitn.it}, \ddag \textcolor{blue}{soumik.bandyopadhyay@unitn.it}, \S \textcolor{blue}{andrea.legramandi@unitn.it}, \# \textcolor{blue}{philipp.hauke@unitn.it}
\end{center}

\section*{Abstract}
{\bf
Complex quantum systems---composed of many, interacting particles---are intrinsically difficult to model. When a quantum many-body system is subject to disorder, it can undergo transitions to regimes with varying non-ergodic and localized behavior, which can significantly reduce the number of relevant basis states.  
It remains an open question whether such transitions are also directly related to an abrupt change in the system's complexity. 
In this work, we study the transition from chaotic to integrable phases in several paradigmatic models, the power-law random banded matrix model, the Rosenzweig--Porter model, and a hybrid SYK+Ising model, comparing three complementary complexity markers---fractal dimension, von Neumann entanglement entropy, and stabilizer Rényi entropy. 
For all three markers, finite-size scaling reveals sharp transitions between high- and low-complexity regimes, which, however, can occur at different critical points. As a consequence, while in the ergodic and localized regimes the markers align, they diverge significantly in the presence of an intermediate fractal phase. 
Additionally, our analysis reveals that the stabilizer Rényi entropy is more sensitive to underlying many-body symmetries, such as fermion parity and time reversal, than the other markers.
As our results show, different markers capture complementary facets of complexity, making it necessary to combine them to obtain a comprehensive diagnosis of phase transitions. 
The divergence between different complexity markers also has significant consequences for the classical simulability of chaotic many-body systems.}

\vspace{10pt}
\noindent\rule{\textwidth}{1pt}
\tableofcontents\thispagestyle{fancy}
\noindent\rule{\textwidth}{1pt}
\vspace{10pt}

\section{Introduction}
\begin{figure}[hbt!]
    \centering
    \includegraphics[width=0.6\columnwidth, clip, trim= 5 175 720 0]{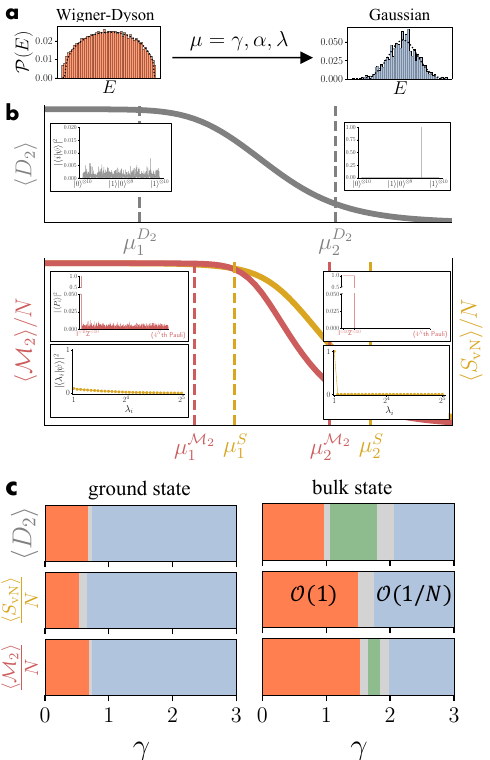}
    \caption{Complexity markers across a transition from a chaotic (ergodic) to an integrable (localized) phase. 
    \textbf{(a)} As the control parameter $\mu$ drives a random matrix model towards localization, the eigenvalue distribution evolves from the Wigner--Dyson semicircle to a Gaussian profile. 
    \textbf{(b)} Comparison of three complexity markers (data for the RP model). A high (low) fractal dimension $D_2$ indicates delocalization (localization) in the computational basis. High  (low) magic $\mathcal{M}_2$ implies support over an extensive set of $4^N$ Pauli strings (concentration on a limited subset). Strong (weak) entanglement $S_\mathrm{vN}$ reflects high (low) Schmidt rank and the presence of many (few) computational basis states in the superposition. 
    While all three markers feature a high-complexity regime, the transition towards intermediate- or low-complexity regimes can happen at distinct points ($\mu_{1,2}^{D_2} \neq \mu_{1,2}^{\mathcal{M}_2} \neq \mu_{1,2}^{S}$). Regimes of different complexity thus overlap but do not coincide.
    \textbf{(c)} The phase transition points in the RP model, obtained through finite-size scaling analysis, differ across the three complexity markers (orange: high [$\mathcal{O}(1)$] complexity; green: intermediate; blue: low complexity [$\mathcal{O}(1/N)$]). Faded regions indicate numerical uncertainty in determining regime boundaries. Notably, the ground state exhibits a sharp transition directly from high to low complexity, bypassing the intermediate regime. In contrast, bulk eigenstates display a well-defined intermediate phase, with a significant variation in transition points across the considered markers.  
    }
    \label{fig: first_figure}
\end{figure}

The interplay between interactions and disorder~\cite{vojta2019disorder} in quantum many-body systems gives rise to complex systems with extremely rich phenomenology, characterized by varying degrees of chaoticity and ergodicity~\cite{gornyi2002quantum,pino2016nonergodic,basko2006metal, pino2016nonergodic, green2015disorder,abanin2019mbl}. 
Although analytical approaches are available in certain limits of ergodic~\cite{Berry:1977wpp,Bohigas:1983er,haake1991quantum,lahoche2020functional} and localized regimes~\cite{Evers_2008,hauke2015mbl,
altman2015universal,goremykina2019analytical}, and while quantum simulations~\cite{hauke2012can,georgescu2014quantum, daley2022practical,Fraxanet2023} have explored various regimes of disordered interacting systems~\cite{jurcevic2014quasiparticle,Schreiber_2015, choi_2016, Smith_2016, Bordia_2017, Luschen_2017, Rispoli_2019, Sauerwein_2023}, their treatment remains a significant challenge both for theory and experiment. 
In order to translate that difficulty into a measurable quantity, various notions have been proposed~\cite{Osborne_2012, orus2019tensor,Baiguera:2025dkc,huang2024non}. 
Notable examples of such  “complexity markers” include entanglement entropy~\cite{horodecki2009quantum} and stabilizer Rényi entropy~\cite{leone2022stabilizer}, 
which have clear implications both for the difficulty of numerically treating~\cite{vidal2003efficient,Eisert2013, orus2014practical,orus2019tensor, 
montangero2018introduction, mora2005algorithmic,
bravyi2019simulation,chernyshev2024quantum, brokemeier2025quantum} the system under study as well as for the quantum resources~\cite{chitambar2019quantum} an experiment needs in order to generate the target state~\cite{joshi2023exploring,howard2017application}. 
Such markers are also deeply connected to the question of what actually defines a quantum system as complex~\cite{liu2018entanglement,baiguera2024complexity, bu2024complexity, haug2024probing, leone2023nonfidelity}.
As a disordered quantum many-body system enters into respectively less ergodic and more localized regimes, 
the number of basis states involved is reduced (see Fig.~\ref{fig: first_figure}\textbf{b}), suggesting intuitively a suppression of quantum complexity. As of yet, it remains, however, an open question how precisely ergodicity transitions are related to changes in complexity markers, and in particular, whether these define qualitatively different complexity regimes. 

In this paper, we map out complexity quantifiers (see Fig.~\ref{fig: first_figure}) in prototypical disordered models that exhibit multiple distinct regimes: the Rosenzweig--Porter (RP) model~\cite{rosenzweig1960repulsion, altland1997perturbation}, the Power-Law Random Banded Matrix (PLRBM) model~\cite{mirlin1996transition}, and the Sachdev--Ye--Kitaev (SYK)~\cite{Sachdev1993,Kitaev2015} model coupled to an Ising chain (see Fig.~\ref{fig:model hamiltonian}). 
The RP and PLRBM are paradigmatic models of chaotic systems that interpolate between ergodic and non-ergodic regimes, and they are central to the understanding of Anderson localization~\cite{anderson1958absence,evers2000fluctuations} and (multi)fractal states~\cite{Evers_2008,kravtsov2015random,Mirlin_2000}. 
The SYK model is a prototypical example of a maximally chaotic quantum many-body system~\cite{Maldacena:2015waa}, which plays a key role in the phenomenological description of non-Fermi liquid behavior~\cite{Sachdev1993,Chowdhury:2021qpy} and in the study of quantum black holes through the holographic principle~\cite{Sachdev:2015efa,Cotler_2017}. 
We scrutinize these models using three different markers of complexity:  
(i) The fractal dimension~\cite{castellani1986multifractal, Halsey_1986,Aoki_1986,kramer1993localization,Cadez_2024}, a key observable for identifying localized regions and mobility edges, quantifies how much a state is spread over a computational basis. 
(ii) The entanglement entropy across half--half bipartitions estimates a state's quantumness in terms of its inability to be expressed as a direct product state~\cite{orus2019tensor}. 
However, not all entangled states are computationally challenging. 
An outstanding example is the class of stabilizer states, 
which contain highly-entangled states but can be tackled efficiently classically thanks to the Gottesman--Knill theorem~\cite{gottesman1998heisenberg,gottesman1998theory}. 
In contrast, non-stabilizer states fall outside this framework and are exponentially hard to represent on a classical computer.
To capture this distinction, we use as the third measure (iii) the non-stabilizerness~\cite{veitch2014resource} (or ``magic'') of a state, quantified through the stabilizer Rényi entropy (SRE)~\cite{leone2022stabilizer}.
While the fractal dimension and entanglement entropy have been instrumental in characterizing non-ergodic quantum systems already for some time~\cite{mirlin1996transition,refael2009criticality,lukin2019probing,Sierant:2024khi},
magic is currently emerging as a powerful tool in various contexts, including the study of ergodicity~\cite{vairogs2024extracting, chen2024magic, zhang2024quantum, turkeshi2023measuring, turkeshi2025pauli, turkeshi2025magic, falcao2025magic}, quantum chaos~\cite{leone2021quantum, goto2022probing, dileepNonstab2025}, and as a sensitive probe for identifying quantum phase transitions~\cite{white2021conformal,oliviero2022magic,tarabunga2023many,falcao2024non, Jasser:2025myz, ding2025evaluating,hoshino2025stabilizer,bera2025non}.

In our work, we find that all three complexity markers show sharp phase transitions, separating regimes of different levels of complexity.  
Notably, the phase transition points are distinct when an extended fractal phase is present, as in the bulk of the RP spectrum. In contrast, when such a phase is absent, as in the ground state of the RP model, all considered markers detect the transition points close to each other, see Fig.~\ref{fig: first_figure}{\bf c}.
Furthermore, an analysis of the SYK$_4$+Ising model reveals that the SRE exhibits a richer structure than the other markers and, in particular, is significantly more sensitive to underlying symmetries of the many-body systems, such as fermion parity and time reversal.
Overall, this comparative analysis suggests that no single marker can independently describe all relevant regimes of the system, which may undergo distinct complexity transitions.
These results not only yield insights into the physics of disordered quantum many-body systems but also help identify the complexity marker best suited to capture the computational hardness across different phases~\cite{yang2017entanglement,ghosh2023complexity,bu2024complexity, chernyshev2024quantum, brokemeier2025quantum}, guiding the selection of appropriate simulation methods.

Finally, our studies complement ongoing research on several complexity markers in ergodic and integrable/localized systems, including Krylov complexity~\cite{Rabinovici:2022beu,basu2024complexity,cohen2024complexity, sahu2024,bhattacharjee2025krylov}, logarithmic multifractality~\cite{chen2024quantum}, operator spreading~\cite{nahum2018operator}, holographic complexity~\cite{alishahiha2015holographic}, 
or quantum-computational complexity~\cite{watrous2008quantum}.

The remainder of the paper is structured as follows. In Sec.~\ref{sec:complexity markers}, we present the complexity markers used to characterize the various quantum phases explored in this study. In Secs.~\ref{sec: RP_model} and \ref{sec: PLRBM model}, we apply these markers to analyze random matrix models, specifically the Rosenzweig–Porter model and the power-law random banded matrix model. In Sec.~\ref{sec: SYK}, we investigate the Sachdev–Ye–Kitaev model interpolated towards the integrable Ising model. Our main findings and possible future research directions are presented in Sec.~\ref{sec: conclusion}. 
In several Appendices, we present technical details on the calculations as well as supplementary findings, e.g., on a mobility edge in the RP model (Appendix~\ref{Appendix: edge mobility}), on transitions in the self-averaging properties of the considered complexity markers (Appendix~\ref{Appendix: self-averaging}), and on the Pauli spectrum of RP and PLRBM models (Appendix~\ref{Appendix: Pauli_spectrum}). 

\section{Complexity markers}\label{sec:complexity markers}

A plethora of different approaches exists to simulate quantum states, each with varying regimes of validity and efficiency, making it essential to scrutinize the complexity of quantum states through multiple markers.
This section introduces the main markers we will employ to analyze how the system evolves from an ergodic to an integrable/localized regime, the fractal dimension, the maximal entanglement entropy, and the stabilizer Rényi entropy (SRE).

\subsection{Fractal dimension}

The fractal dimension can be derived from the inverse participation ratio (IPR), and quantifies the spread of the wavefunction in a given basis, which in our case will be given by the computational basis states $\{ \ket{i} \}^L_{i=1}$, $L$ being the Hilbert space dimension. Fractal dimension and IPR are commonly used markers to investigate many-body localization~\cite{abanin2019mbl,castellani1986multifractal,Mirlin_2000, evers2000fluctuations,Cadez_2024}, as a state that has support on just a few elements of the basis will be in the localized phase, while an ergodic wavefunction will be spread over many states of any generic basis, as shown in Fig.~\ref{fig: first_figure}{\bf b}.
In computational terms, a highly localized state implies a smaller part of the Hilbert space needs to be handled, making it easier to simulate classically~\cite{Prelov_ek_2018,Pietracaprina:2019ihi}. 

For a wavefunction $\ket{\psi}$ expressed in terms of the computational basis with coefficients $\psi(i) = \braket{i|\psi}$, the $q$\textsuperscript{th} IPR is defined as
\begin{equation}\label{eq:IPR}
    I_q(\psi) = \sum^L_{i=1} |\psi(i)|^{2q}, \qquad q>1,
\end{equation}
where $L = 2^N$ is the Hilbert space dimension for a space of $N$ qubits. The fractal dimension $D_q$ is defined from the IPR as~\cite{Halsey_1986,Aoki_1986,kramer1993localization} 
\begin{equation}
    D_q = \lim_{L \to \infty} \frac{\log_L (I_q)}{(1-q)}.
\end{equation}
Typically, one considers $L \gg 1$ and the limit is dropped, such that the fractal dimension is given by 
\begin{equation}
    D_q =\frac{\log_2(I_q)}{N(1-q)}.
\label{Eq: D_q}
\end{equation}

When the wavefunction $\psi$ is confined to a small region of the Hilbert space, the system is considered to be in the localized phase, marked by $D_q=0$ in the thermodynamic limit. For a fully extended state, in contrast, where the wavefunction spreads uniformly over all basis states, one obtains $D_q=1$.
States in the ergodic phase, however, are usually not entirely uniformly distributed. Instead, they can be approximated by Haar-random states, which do not saturate the fractal dimension for finite system sizes.  
For $q=2$, the corresponding fractal dimension is given by~\cite{backer2019multifractal}
\begin{equation}
    D_2^{\rm Haar}= - \frac{1}{N} \log_2\left(\frac{2}{2^N+1}\right) \, .
\end{equation}
For $N\to\infty$, one recovers $D_2^{\rm Haar}\to 1$. 
For non-extended states, as are expected to occur in non-ergodic but delocalized phases, just a fraction of the volume will be occupied, and we will have $D_q < 1$. If $D_q$ is a constant across all values of $q$, the state is called single-fractal (or just fractal), which is the case for the Rosenzweig--Porter model in the non-ergodic extended regime~\cite{kravtsov2015random}. In the general case, $D_q$ is a function of $q$. Such a state is called multi-fractal. An example occurs in the power-law random banded model at the Anderson transition~\cite{mirlin1996transition, evers2000fluctuations}. In this paper, we will focus on $q=2$ only. 

\subsection{Entanglement entropy}

The second marker we consider is the entanglement entropy maximized with respect to all possible equal-sized (half--half) bipartitions of the state in the computational basis. 
The maximal entanglement entropy is a key measure of quantum correlations between subsystems. The presence of entanglement has been identified as one of the first limitations for efficient classical simulations of quantum states~\cite{vidal2003efficient,hastings2007area,eisert2010arealaws}, and entanglement entropy provides an estimate of the bond dimension required in tensor-network methods~\cite {Eisert2013,orus2014practical,montangero2018introduction,orus2019tensor}. Moreover, it is widely used to probe many-body localization and quantum phase transitions~\cite{wu2004quantum, osterloh2002scaling, amico2008entanglement, calabrese2009entanglement, geraedts2016many}.

Given a bipartition of the Hilbert space $ \mathcal{H} = \mathcal{H}_A \otimes \mathcal{H}_B$ and a pure state $\ket{\psi}\in \mathcal{H}$, the von Neumann entropy is defined as $S(\rho_A) = -\mathrm{Tr} (\rho_A \log \rho_A)$, with $\rho_A = \mathrm{Tr} _B \ket{\psi}\bra{\psi}$ the reduced density matrix for subsystem $A$. 
The maximal entanglement entropy is then defined as the von Neumann entropy maximized over all possible half--half bipartitions of the system, 
\begin{equation}
    S_\mathrm{max} = \max_{L_A=\lfloor N/2\rfloor} S(\rho_A)\,.
\end{equation}
For Haar-random states, the entanglement entropy follows the Page curve~\cite{Page:1993df}, which for $L_A \le L_B$ reads
\begin{equation}
    S^{\rm page}_{L_A,L_B}=\sum_{k=L_B+1}^{L_A L_B} \left(\frac{1}{k}\right) -\frac{L_A-1}{2L_B} \,,
\end{equation}
where $L_{A,B}$ is the Hilbert space dimension of $ \mathcal{H}_{A,B}$. In this paper, we work in the convention where logarithms are in base $2$, so that  $S^{\rm page}_{L_A,L_B} / N \approx \frac{1}{2}$ in the large-$N$ limit.

\subsection{Stabilizer Rényi entropy (SRE)}

The last complexity marker we consider is the stabilizer Rényi entropy (SRE). 
The SRE captures how a state is distributed when expressed in the basis of Pauli operators $P \in \mathcal{P}_N$, where $\mathcal{P}_N$ represents the Pauli group for $n$-qubits. Stabilizer states, which are common eigenstates of a maximal set of mutually commuting Pauli operators, exhibit a highly concentrated probability distribution of the Pauli expectation values, confined to a subset of $2^N$ operators, and thus are easily simulatable~\cite{gottesman1997stabilizer,gottesman1998heisenberg,aaronson2004improved,nest2008classical,gidney2021stim,bravyi2019simulation}. In contrast, a Haar-random state has approximately equal weight on all $P$'s~\cite{turkeshi2025pauli}.
Thus, a natural way to quantify the nonstabilizerness of a state $\ket{\psi}$ is by assessing the spread of this distribution using the $q$-th order stabilizer Rényi entropy~\cite{leone2022stabilizer,haug2023stabilizer}~\footnote{Note that, the stabilizer Rényi entropy depends on the computational basis, and hence can be optimized over local rotations~\cite{haug2023quantifying,korbany2025long}.},
\begin{equation}
    \mathcal{M}_q(\psi) = \frac{1}{1-q} \log_2 \sum_{P \in \mathcal{P}_N} \frac{|\langle \psi |P|\psi\rangle|^{2q}}{2^N}\,.
\end{equation}
In this work, we will use the second-order Rényi entropy, $q=2$. 
The SRE vanishes for a stabilizer state and grows as the state exhibits increasing degrees of magic. 
For Haar-random states, we have~\cite{turkeshi2025pauli} (see Appendix~\ref{Appendix: Pauli_&_Haar_random_states} for details),
\begin{equation}
    \mathcal{M}_2^{\rm Haar}=-\log_2\left(\frac{4}{3+2^N}\right) \,. 
\end{equation}
The SRE density for Haar-random states, $\mathcal{M}_2^{\rm Haar}/N$, approaches the maximal value of $1$ in the thermodynamic limit. 
The fractal dimension and entanglement entropy are well-established and commonly used quantities to characterize different regimes of disordered many-body systems. 
Magic offers an alternative perspective on the complexity of states undergoing localization transitions and represents a main focus of this paper.

\begin{figure}[hbt!]
    \centering
    \includegraphics[width=0.65\columnwidth, clip, trim= 0 330 750 0]{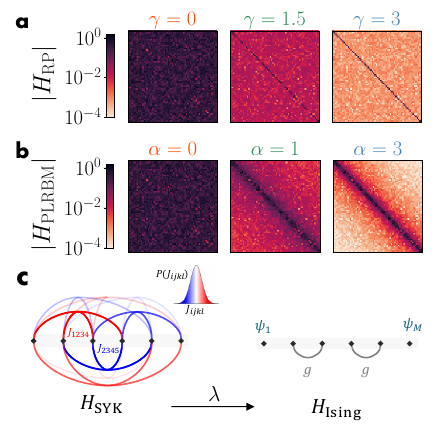}
    \caption{Three model Hamiltonians exhibiting a transition from chaotic (ergodic) to localized phases, controlled by tunable parameters.
    \textbf{(a)} Rosenzweig--Porter (RP) model: A random matrix ensemble where the off-diagonal elements have decreasing weight as the parameter $\gamma$ increases.
    \textbf{(b)} Power-Law Random Banded Matrix (PLRBM) model: Characterized by off-diagonal elements that decay with a power-law determined by the exponent $\alpha$. 
    \textbf{(c)} SYK$_4$+Ising hybrid model: The coupling strength $\lambda$ interpolates between the Sachdev--Ye--Kitaev (SYK$_4$) model with all-to-all random four-body interactions drawn from a Gaussian distribution (left, $\lambda=0$) and an Ising chain, expressed in Majorana fermions (right, $\lambda =1$).}
    \label{fig:model hamiltonian}
\end{figure}

In what follows, we investigate how the three complexity markers discussed above evolve as a system transitions from a chaotic random matrix regime to a localized phase. We consider three representative models: the generalized Rosenzweig–Porter (RP) model (Sec.~\ref{sec: RP_model}), the Power-Law Random Banded Matrix (PLRBM) model (Sec.~\ref{sec: PLRBM model}), and the Sachdev--Ye--Kitaev (SYK$_4$) model coupled to an Ising chain (Sec.~\ref{sec: SYK}). These models are characterized by a single tunable parameter, 
which controls the interpolation between chaotic and localized behavior.
Moreover, the RP and PLRBM models exhibit Anderson criticality and support multifractal eigenstates~\cite{Evers_2008,kravtsov2015random}, providing a fertile ground for analyzing the distinct responses of the complexity markers. 

\section{Rosenzweig--Porter model}~\label{sec: RP_model}

In this section, we study the above complexity markers for the Rosenzweig--Porter (RP) model, both for bulk and ground states. By numerically analyzing varying system sizes, we identify sharp transitions that occur at different points for different markers, and we find distinct behavior in the bulk and ground state.

\subsection{Model}
\label{sec:RP}

The RP model is an ensemble of $L \times L$ random Hermitian matrices defined as~\cite{rosenzweig1960repulsion}
\begin{equation}
    H_{ij} = \delta_{ij} h_i + \frac{1}{L^{\gamma/2}}V_{ij}(1-\delta_{ij})\,.
\label{Eq: RP_model}
\end{equation}
Here, $h_i$ as well as $V_{ij}$ are independent random Gaussian variables (real for $h_i$ and complex for $V_{ij}$) with mean zero and standard deviation $1$.  
At $\gamma=0$, the RP model recovers a Random Matrix Theory (RMT), which in the case of the above choices is given by the Gaussian unitary ensemble (GUE). 
Upon increasing $\gamma$, the model interpolates between this RMT and a diagonal Hamiltonian reached at $\gamma \to \infty$, whose eigenvalues are completely uncorrelated.

The relative simplicity of the RP model enables an analytical treatment of both its spectral properties ~\cite{Kunz_1998,kravtsov2015random} and eigenvector statistics ~\cite{von2019non,Bogomolny_2018_PLRBM,truong2016eigenvectors}, allowing for a comprehensive characterization of the model in the bulk of the energy spectrum.
In the ergodic phase, which spans the range $0 \leq \gamma < 1$, the model behaves like an RMT, with Wigner--Dyson level spacing and Porter--Thomas statistics for the eigenvector probability distribution, signaling quantum chaos and ergodicity in Hilbert space.
At $\gamma = 1$, a transition to non-ergodic behavior occurs, and at $\gamma = 2$ Anderson localization sets in.
In the regime $2<\gamma<\infty $, wavefunctions are localized within a small region of Hilbert space, and the level spacing distribution exhibits Poissonian statistics. 
A key feature of the RP model is the presence of an intermediate regime between the ergodic and localized phases, in the range $1<\gamma<2$, where wavefunctions remain extended but occupy only a fraction of the full Hilbert space, exhibiting fractal behavior that deviates from conventional ergodicity without being fully localized. 
The fractal dimension defined in Eq.~\eqref{Eq: D_q} is a particularly effective marker for these transitions~\cite{kravtsov2015random}, leading to the characterization 
\begin{equation}
\label{eq:D2RP}
   \langle D_q \rangle = \begin{cases}
        1 & \gamma < 1 \, , \\
        2-\gamma & 1 \le \gamma \le 2 \, , \\
        0 & \gamma > 2 \, .
    \end{cases}
\end{equation}
The same phase transitions have been confirmed not only through eigenvalue and eigenstate statistics but also via dynamical properties, such as diffusion characteristics~\cite{monthus2017multifractality, amini2017spread}, as well as survival and return probabilities~\cite{bera2018return, tomasi2019survival}.
Given the presence of different well-established regimes, the RP model serves as an ideal benchmark for evaluating how the behavior of the various complexity markers across different phases, as we discuss in detail in the following section.

\begin{figure*}[hbt!]
    \centering
    \includegraphics[width=\textwidth, clip, trim=0 180 340 0]{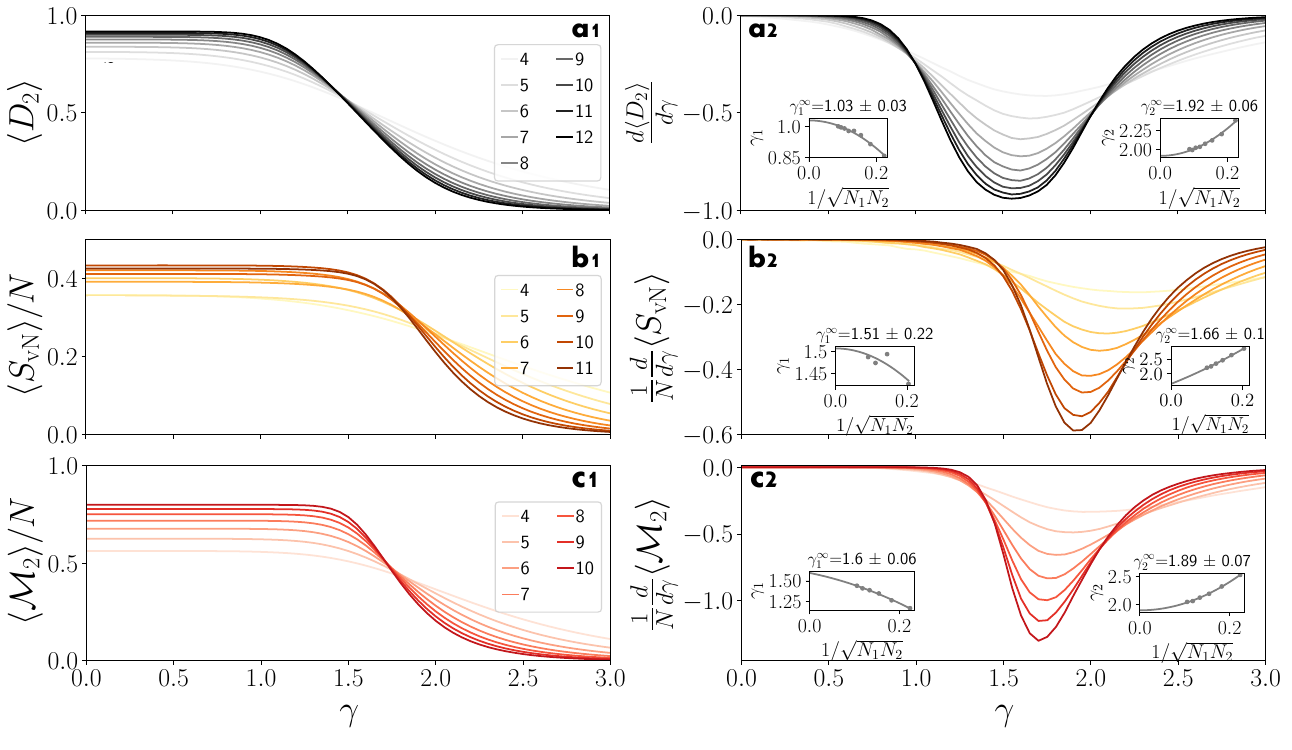}
    \caption{
    (\textbf{a1-c1}) Complexity markers (normalized with respect to $N$), averaged over the central $20\%$ of the bulk eigenstates of the RP model for different system sizes from $N=4$ up to $N=12$, and averaged over samples from $20000$ for $N=4$ to $500$ for $N=12$. All markers plateau at the maximum value in an extended regime at small $\gamma$, overlapping with the model's ergodic regime. The $N$-dependent value matches with predictions from the GUE ensemble that is exactly recovered at $\gamma=0$. The plateau's extent in terms of $\gamma$ depends on the marker: the plateau is largest for $S_{\rm vN}$ and smallest for $D_2$.  
    At large $\gamma$, the model enters a localized regime, characterized by a low value of all markers. 
    The slope at which the model goes from the ergodic to the localized regime is different for all markers, as becomes clear from the derivatives \textbf{(a2-c2)}.   
    The crossings in the derivatives (insets in \textbf{a2-c2}) determine two different transition points $\gamma_{1}$ and $\gamma_2$.
    There are, therefore, three different regimes of high, intermediate, and low complexity, which do not coincide for the different complexity markers used.  
    }
    \label{fig:RP_bulk_all}
\end{figure*}

\subsection{Results}\label{sec: RP_results}

In this section, we comparatively study the behavior of eigenvectors---both ground state and bulk states---undergoing a localization transition using the complexity markers defined in Sec.~\ref{sec:complexity markers}: fractal dimension $\langle D_2 \rangle$, maximum half-chain von-Neumann entanglement entropy $\langle S_{\rm vN} \rangle$, and stabilizer Rényi entropy $\langle \mathcal{M}_2 \rangle$, where the angular brackets denote ensemble averaging. 
We first analyze the bulk eigenstates of the RP model in detail before extending our discussion to the ground state. We focus on the parameter range $\gamma \in [0,3 ]$ for varying system size $N=4$ to $12$ (corresponding to matrix dimension $2^N \times 2^N$), averaged over multiple samples. 

\paragraph*{Bulk state analysis:} To characterize the bulk behavior of the RP model, we compute each complexity marker by averaging over the central 20$\%$ (see Appendix~\ref{Appendix: edge mobility})
~\footnote{We obtain the same result if more than $20\%$ of the bulk eigenstates are choosen instead, see Appendix~\ref{Appendix: edge mobility}} of the ordered eigenstates. At $\gamma = 0$, the system effectively behaves like a Haar-random ensemble, with all three markers saturating their respective analytical values, as described in Sec.~\ref{sec:complexity markers}. In the opposite limit, as $\gamma \to \infty$, the Hamiltonian becomes fully diagonal, leading to eigenstates that are localized in the computational basis. In this regime, both the fractal dimension $D_2$ and the entanglement entropy $S_{\rm vN}$ vanish. Furthermore, since the diagonal Hamiltonian commutes with all $2^N$ Pauli strings composed of only $I$ and $Z$, these are the only ones contributing to the Pauli spectrum. Hence, every eigenstate is a stabilizer state, resulting in vanishing magic. The Pauli spectra for different regimes are shown in the Appendix~\ref{Appendix: Pauli_spectrum}.

For the fractal dimension, two critical points are known within $\gamma \in (0,3)$, corresponding to transitions from ergodic to non-ergodic at $\gamma_1^c=1$ and then to the localized phase at $\gamma_2^c=2$ [Eq.~\eqref{eq:D2RP}]. However, it is not a priori clear that entanglement and magic should exhibit the identical behavior. 
As can be seen in Fig.~\ref{fig:RP_bulk_all}\textbf{b1},\textbf{c1}, at small $\gamma$ they exhibit $\mathcal{O}(1)$ scaling, indicating a high-complexity regime that one may expect in an ergodic phase. At a marker-dependent value $\gamma_1$, the complexity begins to decay, signaling entry into the non-ergodic regime. This decay continues until a second point $\gamma_2$, beyond which the markers eventually stabilize to a $\mathcal{O}(1/N)$ plateau as $N \to \infty$, indicating a low-complexity regime consistent with a large degree of localization. Within the intermediate regime $\gamma_1 < \gamma < \gamma_2$, all markers display fractal-like scaling of the form $\mathcal{O}(N^\delta)$ with $-1 < \delta < 0$.

Notably, the points $\gamma_1$ and $\gamma_2$, and therefore the parameter ranges corresponding to high- and low-complexity behavior, differ significantly between markers. For example, as shown in Fig.~\ref{fig:RP_bulk_all}, the entanglement entropy features the broadest high-complexity plateau (panel \textbf{b1}), while the fractal dimension shows the narrowest (panel \textbf{a1}).
We can make the separation into distinct regimes more precise by examining the first derivative of each marker with respect to $\gamma$. The intersection of these derivatives for different system sizes $N$ provides a reliable signature of emerging non-analytic behavior, indicating the onset of a sharp transition in the thermodynamic limit. We extract the transition points by extrapolating the crossing positions between two consecutive system sizes, $N_1$ and $N_2$, as a function of the inverse geometric mean, $1/\sqrt{N_1 N_2}$~\cite{Pino_2019}. These extrapolations are shown in the insets of Fig.~\ref{fig:RP_bulk_all}{\bf a2}-{\bf c2}.

In the case of $\frac{d \langle D_2 \rangle}{d \gamma}$, the crossing points for consecutive system sizes lie within a relatively narrow range, see the insets of Fig.~\ref{fig:RP_bulk_all}{\bf a2}. By applying a quadratic fit to the crossing locations as a function of $1/\sqrt{N_1 N_2}$ (see Appendix.~\ref{Appendix: scaling of crossing}), we extract the transition points in the thermodynamic limit as $\gamma_1^\infty[D_2] = 1.03 \pm 0.03$ and $\gamma_2^\infty[D_2] = 1.92 \pm 0.06$. These values are in excellent agreement with the analytically predicted transition points from Eq.~\eqref{eq:D2RP}, thereby validating our numerical method. The quoted uncertainties are derived from the covariance matrix of the fit parameters.

In the half-system entanglement entropy, an even--odd behavior is observed in the $\mathcal{O}(1)$ regime: In this regime, the system assumes a volume-law entanglement determined by the subsystem size $\lfloor N/2 \rfloor$, leading to nearly identical half-system entanglement densities for $N$ and $N+1$. 
However, once the system departs from the $\mathcal{O}(1)$ regime, this pairing breaks down, as the volume law no longer holds (see Fig.~\ref{fig:RP_bulk_all}{\bf b1}).
Considering this effect, we use only even $N$ when extracting $\gamma_1^\infty$ via finite-size scaling, while we include all available $N$ for determining $\gamma_2^\infty$. Given the limited number of even-sized systems, the estimate for $\gamma_1^\infty[S_{\rm vN}] = 1.51 \pm 0.22$ has a comparatively larger uncertainty. Even when taking these error bars into account, we have that $\gamma_1^\infty[S_{\rm vN}] > \gamma_1^\infty[D_2]$. Conversely, we find $\gamma_2^\infty[S_{\rm vN}] = 1.66 \pm 0.10$, which is smaller than $\gamma_1^\infty[D_2]$, indicating that the intermediate regime identified by the entanglement entropy is significantly narrower than the non-ergodic fractal regime.

The behavior of magic is smoother across system sizes. The scaling of its crossing points yields a transition point $\gamma_1^\infty[\mathcal{M}_2]=1.60 \pm 0.06$ in the same range as that of the entanglement entropy. However, the transition to the low-complexity $\mathcal{O}(1/N)$-regime occurs at a higher value, $\gamma_2^\infty[\mathcal{M}_2]=1.89 \pm 0.07$, rather more closely aligned with that of the fractal dimension $\gamma_2^\infty[D_2]$. 

Among all the transition points, the most notable contrast lies in $\gamma_1^\infty$, where $S_{\rm vN}$ and $\mathcal{M}_2$ display a new phase transition that occurs significantly later than $\gamma=1$ and that is undetected by $D_2$.
A similar trend is evident in their self-averaging behavior (see Appendix~\ref{Appendix: self-averaging}), which undergoes abrupt changes that reinforce this observation. This behavior may stem from the fact that fractal states, while non-ergodic, can still exhibit high entanglement and high magic~\cite{turkeshi2023measuring}, causing these markers to remain large even after the system departs from full ergodicity. 
Consequently, the extended regime between $\gamma_1^\infty[D_2]$ and $\gamma_1^\infty[\mathcal{M}_2]$ represents a distinct non-ergodic phase, where the states remain highly entangled and highly magical, yet are no longer fully ergodic. This aligns with the intuition that the non-ergodic fractal phase can host highly complex states. 

Finally, we highlight an additional noteworthy observation. As discussed in Appendix~\ref{Appendix: self-averaging}, among the various complexity markers, the SRE exhibits superior self-averaging properties across nearly all regimes, except in the interval $(\gamma_1^\infty[D_2],\gamma_1^\infty[\mathcal{M}_2])$, where the entanglement entropy shows faster convergence.
This enhanced self-averaging behavior makes the SRE a particularly reliable probe for identifying transitions in finite-size systems, where statistical fluctuations are significant. At the same time, the fact that entanglement entropy outperforms SRE in a specific regime underscores the complementary nature of these markers in capturing the structure of complex quantum states.

\begin{figure}[hbt!]
    \centering
    \includegraphics[width=0.65\columnwidth, clip, trim=0 170 610 0]{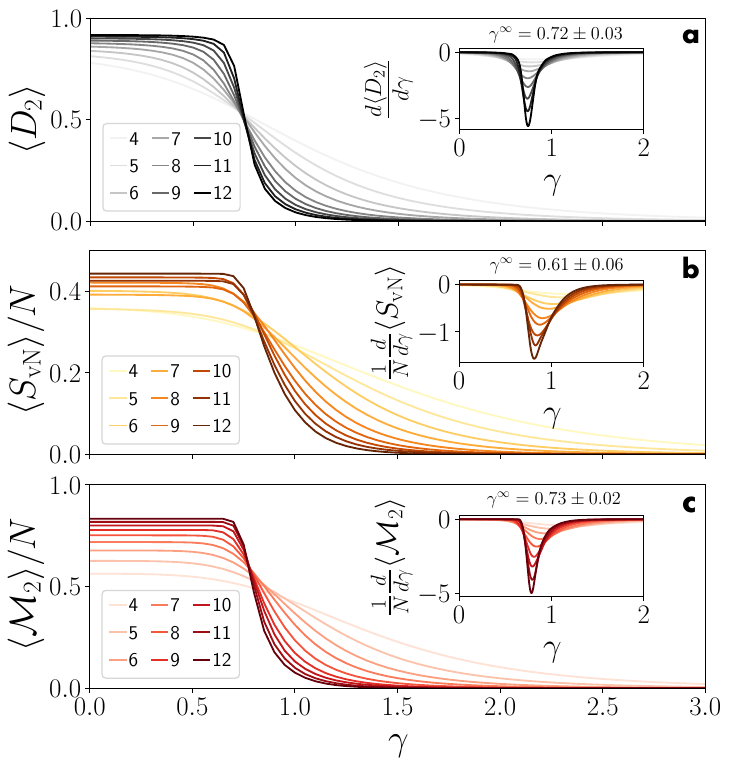}
    \caption{Complexity markers for the ground state of the RP model. Inset: first derivative with respect to $\gamma$. A direct transition from an ergodic to a localized phase is observed. The transition points detected through fractal dimension $D_2$ (\textbf{a}) and SRE density $\mathcal{M}_2/N$ (\textbf{c}) coincide within error bars.  The entanglement entropy $S_\mathrm{vN}$ (\textbf{b}) indicates the transition at a slightly smaller value of $\gamma$, but with a larger error bar due to the even--odd effect.}
    \label{fig: gs_RP_plots}
\end{figure}

\paragraph*{Ground state analysis:}
The analysis of random matrix models is traditionally focused on the properties of the full eigenspectrum.  However, their ground state tends to be more sensitive to ergodicity breaking, giving rise to a mobility edge
~\cite{basko2006metal,sodin2010spectraledgerandomband,
chanda2020many, guo2021observation, geissler2020mobility, wei2020characterization, yousefjani2023mobility} and motivating a dedicated analysis.  

As shown in Fig.~\ref{fig: gs_RP_plots}, again all complexity markers approach their Haar-random values in the small-$\gamma$ regime, with complexity scaling as $\mathcal{O}(1)$. However, they begin to deviate from this value at a smaller $\gamma$ than their bulk counterparts. Such an earlier transition point in the ground state compared to the bulk indicates the presence of a mobility edge (see Appendix~\ref{Appendix: edge mobility}). 
Furthermore, all markers transition more rapidly to a $\mathcal{O}(1/N)$ scaling than was the case in the bulk. As a result, in the large-$N$ limit the intermediate region with scaling $\mathcal{O}(1/N^\delta)$ appears to vanish entirely, giving rise to a sharp, first-order-like transition directly from $\mathcal{O}(1)$ to $\mathcal{O}(1/N)$ scaling. These abrupt transitions are identified by extrema in the derivatives of the complexity markers~\footnote{Similar to bulk states, the derivatives of the ground state's complexity marker for systems of size $N$ and $N+1$ intersect at two distinct points. However, after finite-size scaling, these crossings converge to a single transition point, which is consistently identified by the extrema of the derivative.}, signaling non-analytic behavior in the thermodynamic limit (see insets of Fig.~\ref{fig: gs_RP_plots}). After finite-size scaling, we find that both the fractal dimension and magic detect the ground-state transition at around the same value, $\gamma^\infty[D_2] = 0.72\pm 0.03$ and $\gamma^\infty[\mathcal{M}_2]=0.73\pm 0.02$, respectively. 
In contrast, for entanglement it appears somewhat earlier, at $\gamma^\infty[S_{\rm vN}] = 0.61\pm 0.06$, but again with a larger error bar.
Given the relative vicinity between the different transition points, we cannot exclude the possibility that they identify the same phase transition. 
Above, we found that the transitions in $S_{\rm vN}$ and $\mathcal{M}_2$ occurred well within the fractal regime. One may thus conjecture that all transition points agree when an intermediate fractal regime is missing.

\section{Power-Law Random Banded Matrix model}\label{sec: PLRBM model}
To further deepen our understanding of the complexity transitions in random matrix models, in this section we examine another paradigmatic model—the Power-Law Random Banded Matrix (PLRBM) model. 

\subsection{Model}
The PLRBM model~\cite{mirlin1996transition} is conceptually similar to the RP model, as it is a random matrix model with suppressed off-diagonal elements. Nevertheless, its physics and phase diagram have distinct features from the RP model. The PLRBM model is a random matrix ensemble whose elements are given by 
\begin{equation}
    H_{ij} = G_{ij} a(|i-j|)\,.
\end{equation}
In our case, $G$ is chosen as a GUE matrix. The coefficients $a(r)$ implement a polynomial decay of the off-diagonal elements beyond a certain range, 
\begin{equation}
  a(r) =
    \begin{cases}
      1, & r < b\\
      (r/b)^{-\alpha} & r \geq b,
    \end{cases}    
\label{Eq: PLRBM Hamiltonian}
\end{equation}
where $b$ is called the bandwidth parameter and $\alpha$ is the decay exponent. In this work, we set $b=1$, i.e., only terms adjacent to the diagonal remain intact, and $\alpha$ remains the single tunable parameter. 

The model is known to behave like a random matrix theory for $\alpha < \frac{1}{2}$~\cite{mirlin1996transition,xu2023non}. In the regime $\frac{1}{2}<\alpha < 1$, the eigenvectors depart from the Porter--Thomas distribution and follow a generalized hyperbolic distribution~\cite{Bogomolny_2018_PLRBM}, signaling the presence of a weakly-ergodic regime distinct from that of the RP model. 
A recent study on the variable-range SYK$_2$ model reveals clear signatures of this non-ergodic transition at $\alpha = \frac{1}{2}$ in the many-body spectral statistics~\cite{Legramandi:2024kcv}.
There is an Anderson critical point at $\alpha = 1$~\cite{Varga_2000,mirlin1996transition}, where the eigenvectors display proper multifractal behavior that has been intensively studied in the literature~\cite{evers2000fluctuations,Mirlin_2000,Evers_2008}. In the range $1 < \alpha < \frac{3}{2}$, the PLRBM enters a localized yet superdiffusive phase, while it becomes fully localized for $\alpha > \frac{3}{2}$. 

The differences between the PLRBM and RP model provide a valuable framework for a comparative analysis of how the complexity markers behave in the different regimes, as we will see in what follows. In this section, we focus on the bulk, again considering the central $20\%$ of the energy eigenstates.

\subsection{Results}\label{sec: PLRBM_Results}

\begin{figure*}[hbt!]
    \centering
    \includegraphics[width=\textwidth,clip, trim=0 180 340 0]{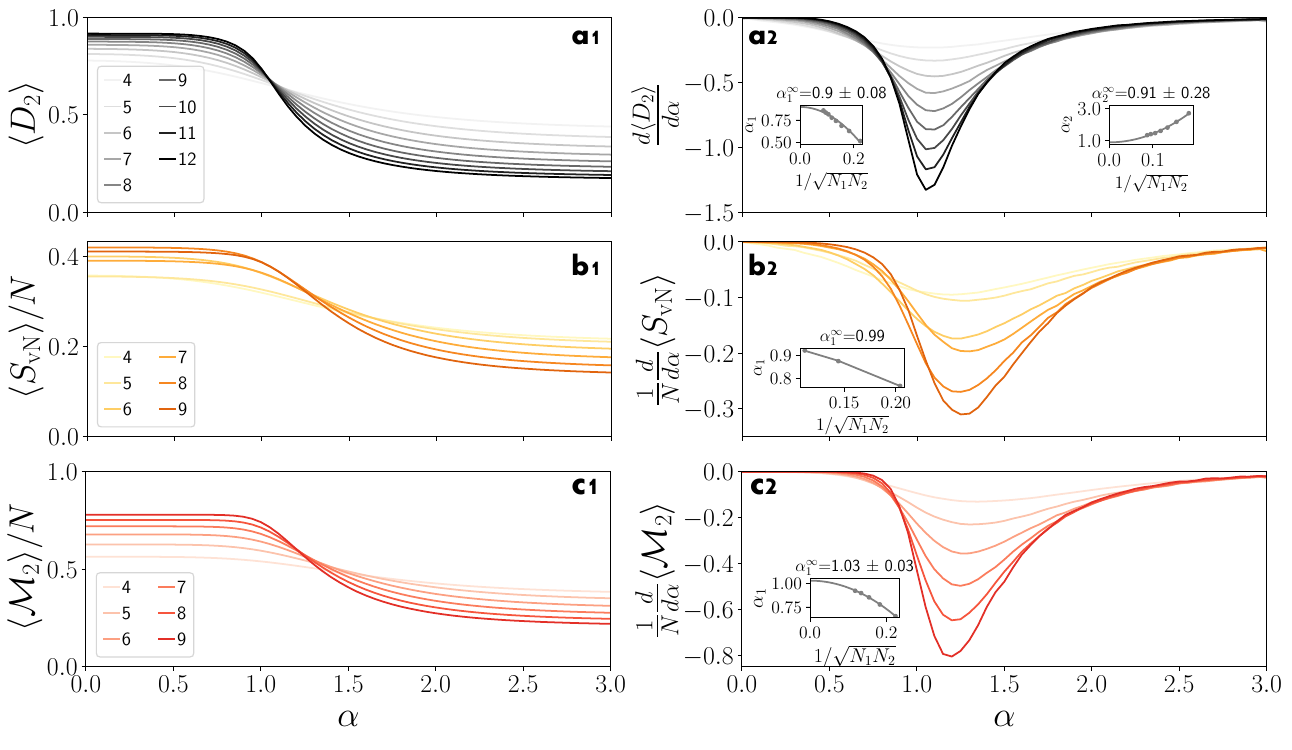}
    \caption{(\textbf{a1-c1}) Complexity markers (normalized with respect to $N$), averaged over the central $20\%$ of bulk eigenstates of the PLRBM model for $N=4$ up to $N=9$ ($N=12$ for the fractal dimension) averaged over different sample sizes of $35000$ to $3000$, respectively. Throughout an extended regime at small $\alpha$, all markers plateau at the value matching with predictions from the GUE ensemble (recovered by the model at $\alpha=0$). At large $\alpha$, the model transitions to a localized regime, characterized by a lower value of all the markers. The transition point can be deduced from the derivatives of the complexity markers (\textbf{a2-c2}). The crossings in the derivatives (insets \textbf{a2-c2}) effectively find only one distinct transition point. Therefore, in contrast to the RP model, there are only 2 different regimes of high and low complexity that coincide with the ergodic
    and localized phases, respectively.}
    \label{fig: RBM_results_bulk}
\end{figure*}

The numerical results for the three complexity markers as a function of $\alpha \in [0,3]$ are plotted in Fig.~\ref{fig: RBM_results_bulk}. Note that compared to the RP model, the PLRBM model has more significant fluctuations across realizations due to the persistence of some long-range correlations, which do not vanish in the large $N$ limit (see Fig.~\ref{fig:model hamiltonian}\textbf{b}). For instance, while $1000$ samples suffice for the RP model, we use $10000$ samples for the PLRBM at $N=8$ (see Appendix~\ref{Appendix: number of samples}) when computing the SRE---the most self-averaging of the three markers---yet noticeable fluctuations persist.

At $\alpha=0$, the state behaves again equal to a Haar-random state and reaches the maximal value for all complexity markers, similar to the RP model. This value extends up to around $\alpha=1$, creating a plateau with value $\mathcal{O}(1)$ and indicating the extended ergodic regime of the model. Depending on the marker, the complexity begins to decay at a specific value $\alpha_1$, signaling the entry of the system into the non-ergodic regime. 
In the limit for $\alpha \to \infty$, the PLRBM model becomes a tridiagonal matrix. Hence, it does not commute with all the $2^N$ diagonal Pauli strings and, consequently, the magic is nonzero (see Appendix~\ref{Appendix: Pauli_spectrum}). For the same reason, the eigenstates cannot completely localize in the computational basis. Indeed, in Fig.~\ref{fig: RBM_results_bulk} we see that for large values of $\alpha$ the markers do not vanish and still show a finite $N$-dependence. As before, we identify the phase transition points for each complexity marker by considering the positions of the crossing points of the first derivatives in the thermodynamic limit. 

For the fractal dimension, we find two crossing points $\alpha_1^\infty[D_2] = 0.90 \pm 0.08$ and $\alpha_2^\infty[D_2] = 0.91 \pm 0.28$, indicating that, in the large $N$ limit, they will collapse to the same value. 
The curves for entanglement entropy and magic at different system sizes cross just at one transition point $\alpha_1^\infty[S_{\rm vN}] = 0.99$~\footnote{For $S_{\rm vN}$ the pairing behavior between even and odd numbers of qubits for the entanglement entropy is still visible in the $\mathcal{O}(1)$ regime. This behavior forces us to restrict our analysis to, e.g., the even $N$ curves in order to locate the transition. We compute some extra values of $S_{\rm vN}$ for $N=10$ around the transition to obtain an additional crossing point. Thus, we obtain three crossing points---for instance, between (4,6),(6,8),(8,10). While this data allows us to fit a quadratic function, which can retrieve the Anderson-localization transition at $\alpha=1$, the limited number of points is not sufficient to estimate an error bar.} and $\alpha_1^\infty[\mathcal{M}_2] = 1.03 \pm 0.03$
(There might be a second crossing point that remains undetected due to the more sizable fluctuations in the localized regime for large $\alpha$.) 
Thus, in contrast to the RP model, our analysis does not detect an intermediate regime that is non-ergodic but delocalized. Rather, the available data suggests a direct transition from the delocalized to the localized phase close to the Anderson critical point at $\alpha = 1$.

Although the analysis of the PLRBM model is restricted to smaller system sizes and has more sizable fluctuations, it supports what we conjectured for the RP model: the complexity markers assume distinct values across the ergodic-to-non-ergodic transition only in the presence of a genuine fractal (or multifractal) phase. Indeed, within error bars, in the PLRBM model, the transition points identified by all three complexity markers coincide (or almost coincide) with the Anderson critical point, the only transition captured by the fractal dimension in the thermodynamic limit. 
We do not report here the results for the ground state behavior of the PLRBM model since they qualitatively resemble that of the RP model considered in Sec.~\ref{sec: RP_results}: the transitions occur earlier than in the bulk~\cite{buijsman2025power}, yet no intermediate phase is observed. This is consistent with expectations, as the markers used do not reveal a clear extended non-ergodic regime even in the bulk of the PLRBM model.

\section{SYK$_4$+Ising model} \label{sec: SYK}

In this section, we consider a quantum many-body model subject to a chaotic--integrable transition to contrast with the random-matrix behavior. For this, we choose a hybrid model consisting of the SYK model interpolating towards the Ising model. The SYK model, originally introduced as a phenomenological model for non-Fermi liquids~\cite{Sachdev1993}, is a quantum mechanical system consisting of $M$ Majorana fermions with random $q$-body interactions. It has accumulated significant attention due to the combination of strong interactions and solvability in the large-$M$ limit~\cite{Kitaev2015,Maldacena:2016hyu}, making it a paradigmatic system to study quantum chaos beyond RMT. Moreover, randomly interacting many-body systems, like the SYK model, are known to behave drastically different from RMT near the edge of their energy spectrum~\cite{Quantum_chaos_on_edge}. This makes it particularly interesting to compare the complexity markers for the ground state of this model to those of the RP model of Sec.~\ref{sec: RP_results}. Further, the SYK model is a prototypical system in the study of black holes and quantum gravity~\cite{Sachdev:2015efa,Cotler_2017} through the AdS/CFT correspondence~\cite{Maldacena:1998im,Witten:1998qj}. Indeed, at low energies, it exhibits an emergent conformal symmetry and is effectively described by a Schwarzian action~\cite{Maldacena:2016hyu}, mirroring the behavior of Jackiw--Teitelboim gravity~\cite{Maldacena:2016upp}. Since the duality becomes manifest in the infrared, it is of particular interest to understand the ground-state properties of the model in order to assess the stability of the holographic phase under perturbations.

\subsection{Model}
The system under study is composed of $M$ interacting Majorana fermions, described by operators $\psi_i$ fulfilling the commutation relations $\left\{ \psi_i, \psi_j \right\} = \delta_{ij}$. The Hilbert space of the model is isomorphic to that of $N=M/2$ qubits. The interactions between the fermions are chosen to interpolate between a chaotic Hamiltonian at $\lambda=0$, provided by the SYK model, and a fully localized, integrable Ising chain reached at $\lambda=1$, 
\begin{equation}
    H = (1-\lambda) H_{\text{SYK$_4$}} + \lambda H_{\rm Ising}\,.
\label{Eq: SYK$_4$+Ising}
\end{equation}

The SYK$_q$ model is a system with $M$ all-to-all interacting Majorana fermions with random $q$-body interactions, (see Fig.~\ref{fig:model hamiltonian}\textbf{c} where $q=4$) , defined as
\begin{equation}
H_\text{SYK$_q$} = (i)^{q/2} \sum_{1 \leq i_1 < \dots < i_q \leq M} J_{i_1 \dots i_q} \psi_{i_1} \dots \psi_{i_q}\,,
\end{equation}
where the couplings $J_{i_1 \dots i_q}$ are independent and identically distributed random variables drawn from a Gaussian distribution with zero mean and variance given by $\langle J_{i_1 \dots i_q}^2 \rangle = J^2 (q-1)!/M^{q-1}$. In this article, we work with $q=4$.

The Ising Hamiltonian is given by an anti-ferromagnetic ZZ spin chain without magnetic field, $ H_{\rm Ising} = \sum_{i=1}^{N-1} g \sigma^z_i \sigma^z_{i+1}$. It can be written in terms of Majorana operators using the Jordan--Wigner transformation, which maps each Majorana operator to a Pauli string consisting of $N$ Pauli matrices~\cite{Maj_to_Pauli},
\begin{align}
    \psi_{2i-1} &= \sigma^x_1 \otimes \ldots \otimes \sigma^x_{i-1} \otimes \sigma^z_i \otimes \mathbbm{1}_{i+1} \otimes \ldots \otimes \mathbbm{1}_{N}, \\
    \psi_{2i} &= \sigma^x_1 \otimes \ldots \otimes \sigma^x_{i-1} \otimes \sigma^y_i \otimes \mathbbm{1}_{i+1} \otimes \ldots \otimes \mathbbm{1}_{N}\,.
\end{align}
Using
\begin{equation}
    \psi_{2i}\psi_{2i + 1} = \mathbbm{1}_1 \otimes \ldots \otimes \mathbbm{1}_{i-1} \otimes -i \sigma^z_i \otimes \sigma^z_{i+1} \otimes \mathbbm{1}_{i+2} \otimes \ldots \mathbbm{1}_N 
\end{equation} 
we obtain 
\begin{equation}
    H_{\rm Ising} = i \sum_{i=1}^{M/2-1} g \psi_{2i} \psi_{2i+1}\,,
\label{Eq: Ising_Majorana}
\end{equation}
where only interactions between even-odd Majorana fermions remain, see Fig.~\ref{fig:model hamiltonian}\textbf{c}.

The Ising contribution to the model in Eq.~\eqref{Eq: Ising_Majorana} introduces purely neighboring and equal-strength interactions between the Majorana fermions. Thus, an increasing $\lambda$ gradually suppresses the disordered all-to-all interactions characteristic of the ergodic SYK$_4$ regime. The effect is reminiscent of the suppression of off-diagonal terms in the RP model in Eq.~\eqref{Eq: RP_model} considered above. 

The symmetries of this model play an important role in understanding the behavior of complexity markers, as will be discussed in detail in the following section. Both the SYK$_4$ and Ising model Hamiltonian commute with the fermion parity operator
\begin{equation}
    (-1)^\mathrm{F} = (i)^M \psi_1 \dots \psi_M = (\sigma^x)^{\otimes M} \, .
    \label{eq:ferm_parity}
\end{equation}
In addition, $H_{{\rm SYK}_4}$ commutes with the anti-unitary time reversal operator $\mathcal{T}$. The interplay between $(-1)^\mathrm{F}$ and $\mathcal{T}$ determines the RMT ensemble and degeneracy of the ground state of the SYK model~\cite{Li:2017hdt}: for $M \bmod 4 = 2$ the ensemble corresponds to GUE, for $M \bmod 8 = 0$ the corresponding ensemble is GOE, and for $M \bmod 8 = 4$ it is GSE \footnote{The Gaussian symplectic ensemble describes the random $L \times L$ Hermitian quaternionic matrices. The numerical results were obtained by representing the GSE matrices as $2L \times 2L$ complex matrices. For consistency, when comparing the SYK$_4$ results falling in the GSE symmetry class with the corresponding Haar-random value, we take an ensemble of dimension $L^{N-1}$ such that the Hilbert space dimension of SYK$_4$ matches with the dimension of the complex GSE representation.}. In contrast, $H_\text{Ising}$ as given in Eq.~\eqref{Eq: Ising_Majorana} anticommutes with $\mathcal{T}$ due to the presence of the imaginary unit in the Hamiltonian, lifting for $\lambda \neq 0$ the symmetry class to GUE for all $M$.

\subsection{Results}

Similarly to the RP and PLRBM models, we carry out a comparative analysis of three complexity markers for the SYK$_4$
+ Ising model. 
Additionally, we present a more detailed study of magic across different $M$, which turns out to be sensitive to the model's symmetries introduced above. 
To be consistent with the rest of the paper, we consider the number of qubits $N = M/2$ instead of the number of Majorana fermions when calculating the density of the markers.

\begin{figure}[hbt!]
     \centering
    \includegraphics[width=0.65\columnwidth, clip, trim=0 130 600 0]{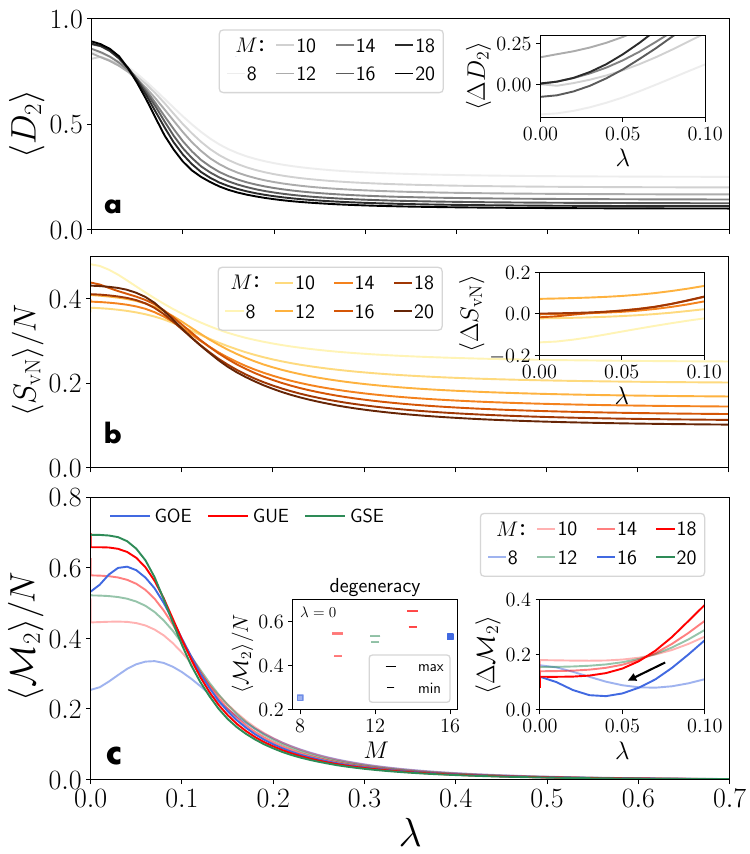}
    \caption{Ensemble averaged complexity markers for the groundstate of the SYK$_4$ + Ising model for different number of Majorana fermions $M$. (\textbf{a}) The fractal dimension $D_2$ starts to decrease for any non-zero $\lambda$ and approaches the asymptotic value of the Ising model relatively quickly. Inset: Difference of $D_2$ to the Haar-random value, indicating that for sufficiently large $M$, $D_2$ saturates this value at $\lambda=0$. (\textbf{b}) The $S_{\rm vN}$-density shows a small remnant of a plateau around the SYK$_4$ limit ($\lambda=0$). 
    Inset: for large $M$ also the $S_{\rm vN}$-density approaches the Haar-random value at small $\lambda$. (\textbf{c}) For $\lambda \gtrsim 0.5$, the average magic remains zero, while it reaches its maximum near $\lambda \sim 0$, except for states corresponding to the GOE ensemble (which reflects the symmetry of the SYK model at $\lambda = 0$), where the peak occurs away from zero. Left inset: At $\lambda = 0$, the ground states corresponding to the GUE and GSE ensembles are degenerate, where a range of SRE values is possible, however, as soon as  $\lambda$ deviates from zero, it chooses the minimum. Right inset: The difference to the Haar-random value is larger than for $D_2$ and $S_{\rm vN}$.}
    \label{fig:SYK_results}
\end{figure}

In the integrable Ising model at $\lambda = 1$, the ground state is doubly degenerate. However, introducing a small perturbation from the SYK Hamiltonian, $\lambda \lesssim 1$, lifts this degeneracy. Due to the fermion parity symmetry of the full Hamiltonian, the system selects a specific ground state: an equal superposition of the two antiferromagnetic spin configurations, with a relative phase of $\pm 1$.
As a result, in this limit, the fractal dimension and maximal von-Neumann entropy density approach the values of $1/N$. In contrast, the magic tends to zero since this superposition is a stabilizer state (see Appendix~\ref{Appendix: SYK_magic}).  
All considered markers approach the values of the $\lambda \sim 1$ limit relatively rapidly without appreciable changes beyond $\lambda \gtrsim 0.5$, see Fig.~\ref{fig:SYK_results}.

In the chaotic regime for $\lambda \gtrsim 0$, unlike in the RP ground state, we cannot a priori know if the complexity markers saturate the Haar-random values since we are now considering a many-body system. The deviation from the Haar-random value is quantified by
\begin{equation}
    \langle \Delta \mathcal{C} \rangle = \frac{\langle \mathcal{C}_{\rm Haar} \rangle}{N} - \frac{\langle \mathcal{C} \rangle}{N},
\label{Eq: C_difference}
\end{equation}
where $\mathcal{C}$ denotes one of the three markers $D_2, S_{\rm vN}$, or $\mathcal{M}_2$. As shown in the insets of Fig.~\ref{fig:SYK_results}, for sufficiently large $M$, $\langle \Delta D_2 \rangle$ and $\langle \Delta S_{\rm vN} \rangle$ tend to zero as $\lambda \to 0$, whereas $\langle \Delta \mathcal{M}_2\rangle$ is strictly larger than zero.
This behavior can be understood by the fact that, similarly to what we have seen for $\lambda \sim 1$, the ground state for small $\lambda$ has fixed fermion parity, which means that only Pauli strings composed of the product of an even number of Majorana operators (those that preserve the symmetry) contribute to a non-zero \(\mathcal{M}_2\). Since these comprise only half of all possible Pauli strings, the overall $\mathcal{M}_2$ is smaller than the Haar-random value.
The situation is, however, a bit more subtle in the strict $\lambda=0$ case since in pure SYK$_4$ the ground state is not always unique. For $M = 0 \bmod 8$ (GOE), there is just one ground state with fixed fermion parity. In contrast, for GUE and GSE, $M \neq 0 \bmod 8$, the presence of degeneracy allows for the superposition of the two ground states with opposite fermion parity, which can enhance the SRE.  We numerically determine the maximum and minimum values of this linear combination \footnote{Given the two ground states $\rm GS_1$ and $\rm GS_2$ with opposite fermion parity and a generic linear combination
\begin{equation}
    \ket{\psi} = \sin\theta \ket{\rm GS_1} + e^{i\phi}\cos\theta\ket{\rm GS_2}, 
\end{equation}
we numerically optimize \(\langle \mathcal{M}_2(\psi) \rangle\) over a \(200 \times 200\) grid in the \((\theta, \phi)\) plane to determine the maximum and minimum values.}, as shown in the left inset of Fig.~\ref{fig:SYK_results}. Notably, the minima coincide with the values shown in the main plot (\textbf{c}) at small but non-zero $\lambda$, confirming that the Ising Hamiltonian breaks the degeneracy and selects one specific parity.

Moreover, the SRE exhibits a peculiar $M$-dependent behavior at small $\lambda$, 
which can be attributed to the characteristics of the underlying RMT ensembles~\cite{Cotler_2017,Kanazawa:2017dpd}. Specifically, when \(M \bmod 8 = 0\), \(\mathcal{M}_2\) begins at a lower value, and then rises to a peak at some small nonzero \(\lambda\), followed by a monotonic decrease. This behavior can be comprehended in the following way: for finite-size systems, the GOE ensemble yields lower magic than GUE or GSE for Haar-random states (see Eq.~\eqref{Eq: Magic_RMT_analytical} and Fig.~\ref{fig: analytical_gaussian_magic} in Appendix~\ref{Appendix: Pauli_&_Haar_random_states}). As contributions from the Ising Hamiltonian kick in at $\lambda\neq0$, the symmetry class is lifted from GOE to GUE, leading to a rise of the value for the SRE in the case of \(M \bmod 8 = 0\). In contrast, for \(M \bmod 4 = 2\) and \(M \bmod 8 = 4\), the SRE starts at its maximal value and the addition of the Ising contribution only decreases it. 

As we move further away from $\lambda\sim 0$, all three markers start deviating from their ergodic value without exhibiting a long plateau as in the RP model---indicating that the ground state is rather susceptible to perturbations by the Ising contribution (see Fig.~\ref{fig:SYK_results}), unlike the $\lambda\gtrsim 0.5$ regime. Such fragility of the SYK$_4$ ground state resembles recent observations in a related model interpolating between SYK$_4$ and SYK$_2$~\cite{Jasser:2025myz}, which implements a crossover from many-body chaos to single-body chaos, while always maintaining volume-law magic in the ground state. 
In our model, the perturbed Ising ground state is a stabilizer state; therefore, the transition in magic appears sharper, reminiscent of the transition observed in the RP ground state (Fig.~\ref{fig: gs_RP_plots}).

We expect the ensemble-dependent features discussed above to disappear in the \(M \to \infty\) limit. Indeed, the \(M \bmod 8 = 0\) peak seems to progressively move to smaller values of $\lambda$ and \(\Delta \mathcal{M}_2 \) progressively decreases if we compare the same ensembles. Nonetheless, such finite-size effects are particularly relevant for characterizing realistic many-body chaotic systems beyond random matrix universality~\cite{Quantum_chaos_on_edge,Legramandi:2024ljn}. 
The analysis highlights the utility of the SRE as a diagnostic tool due to its sensitivity to the model's symmetry structure, distinguishing it from other markers such as the fractal dimension and entanglement entropy.

\section{Conclusion}\label{sec: conclusion}

In this work, we have conducted a comparative study of various complexity markers across families of Hamiltonians that interpolate between ergodic, extended non-ergodic, and localized phases. When a fractal region in the parameter space is present, such as for states in the bulk of the Rosenzweig--Porter model spectrum, the different markers identify high and low complexity regimes in overlapping but distinct parameter regions. 
Both entanglement entropy and stabilizer Rényi entropy undergo a transition within the fractal phase that is not detected by the fractal
dimension, indicating the existence of a non-ergodic yet complex regime with respect to these two markers. 
Moreover, the stabilizer Rényi entropy reveals an additional extended intermediate regime with non-maximal magic, before eventually dropping to low complexity in correspondence to the localization point. 
The presence of an extended fractal region seems to be crucial to the appearance of new complexity regimes: when such a region is absent, as in the ground state of the Rosenzweig--Porter model and the bulk of the power-law random banded model, the transition points identified by all complexity markers remain consistent. 
The behavior of the stabilizer Rényi entropy suggests that it can uncover subtle features in the quantum phase diagram that might otherwise go unnoticed using conventional localization markers.
It would be interesting to corroborate these findings through analytical studies, e.g., of the Rosenzweig--Porter model akin to those employed for fractal dimensions~\cite{kravtsov2015random}.

Moreover, our analysis of the ergodic to integrable transition of the SYK$_4$ + Ising model suggests that stabilizer Rényi entropy may be a more sensitive marker to the underlying symmetries of the model, as indicated by its non-monotonic behavior as compared to the fractal dimension and the entanglement entropy. 
This observation motivates a more systematic investigation of the relationship between non-stabilizerness and symmetry classes in SYK-type models, extending beyond the cases considered in Refs.~\cite{Li:2017hdt,Kanazawa:2017dpd}. Additionally, it would be valuable to study other quadratic perturbations of the SYK model to determine whether the complexity markers studied here can consistently capture the observed fragility of ground-state ergodicity. This is especially relevant for assessing the stability of the holographic phase, which has been argued to remain robust under deformations only within a vanishingly small region in the thermodynamic limit~\cite{Lunkin_2018}

Investigating the presence of non-ergodic but highly complex regimes would not only shed light on the structure of the quantum phase diagram but also help assess the sensitivity of non-stabilizer magic to non-ergodic behavior and complexity transitions.
This could be pursued by analyzing many-body systems undergoing a many-body localization transition, where multifractal phases are expected to arise, such as disordered spin systems~\cite{Torres_Herrera_2017,mace2019multifractal,mace2019multifractal} and the Bose--Hubbard model~\cite{lindinger2019many, geissler2020mobility}. 
Localization also plays a significant role in quantum optimization~\cite{altshuler2010anderson, knysh2010relevance, pino2018quantum, cao2024exploiting, stark2011localization,cao2024exploiting}, whose target problems are often described by spin-glass models~\cite{panchenko2013sherrington, venturelli2015quantum,hauke2020perspectives}.
In these contexts, our work offers a promising perspective for probing complex characteristics of physical systems approaching the many-body localization transition.

The presence of distinct crossover values at finite system sizes is crucial for both classical and quantum simulations, as the feasible techniques are often constrained by available quantum resources. For instance, in noisy intermediate-scale quantum (NISQ) devices, the bottleneck is the ability to build up entanglement across the device~\cite{preskill2012quantum}, while in tensor networks, the reachable bond dimension constrains the accessible entanglement entropy~\cite{Eisert2013, orus2014practical, orus2019tensor, montangero2018introduction}. In contrast, the efficiency of classical calculations using the Gottesmann--Knill framework~\cite{gottesman1997stabilizer,gottesman1998heisenberg,aaronson2004improved,nest2008classical,gidney2021stim} as well as the cost of broad classes of fault-tolerant quantum computers~\cite{google2023suppressing,katabarwa2024early} is determined by the number of $T$-gates~\cite{bravyi2019simulation, ogorman2017quantum}, which is related to the magic in a quantum state~\cite{howard2017application}. In this context, our analysis could guide the identification of the most efficient simulation strategies for each regime.

\section*{Acknowledgments}
We thank Julius Mildenberger for valuable discussions regarding the calculation of magic and Xhek Turkeshi for discussions on the analytical expression of Haar-random states.

\paragraph{Funding information}
This project has received funding by the European Union under Horizon Europe Programme - Grant Agreement 101080086 - NeQST and under NextGenerationEU via the ICSC – Centro Nazionale di Ricerca in HPC, Big Data and Quantum Computing. 
This work was supported by the Swiss State Secretariat for Education,
Research and Innovation (SERI) under contract number UeMO19-5.1.
This project has received funding from 
the European Union - Next Generation EU, Mission 4 Component 2 - CUP E53D23002240006 and
the Italian Ministry of University and Research (MUR) through the FARE grant for the project DAVNE (Grant R20PEX7Y3A), was supported by the Provincia Autonoma di Trento, and by Q@TN, the joint lab between the University of Trento, FBK-Fondazione Bruno Kessler, INFN-National Institute for Nuclear Physics and CNR-National Research Council.
S.B.\ acknowledges CINECA for use of HPC resources under Italian SuperComputing Resource Allocation– ISCRA Class C
Project No. DeepSYK - HP10CAD1L3. Views and opinions expressed are however those of the author(s) only and do not necessarily reflect those of the European Union or the European Commission. Neither the European Union nor the granting authority can be held responsible for them.

\begin{appendix}

\section{Mobility edge in the energy spectrum}\label{Appendix: edge mobility}

The transition to the localized phase can be energy-dependent, a phenomenon referred to as a mobility edge~\cite{basko2006metal,chanda2020many, guo2021observation, geissler2020mobility, wei2020characterization, yousefjani2023mobility}. For instance, low-energy states can become localized at lower values of the parameter regime compared to states near the middle of the spectrum. Random banded models are known to exhibit such a mobility edge~\cite{sodin2010spectraledgerandomband}. A similar trend is observed in our numerical analysis of the RP model (see Fig.~\ref{fig: first_figure}\textbf{c}). As a result, for certain values of the control parameter, low-energy states may already be localized while high-energy states remain ergodic.
Given this sensitivity to the energy window, it is crucial to investigate how different complexity markers respond to variations in the chosen spectral region.

\begin{figure}[hbt!]
    \centering
    \includegraphics[width=0.7\columnwidth, clip, trim=0 50 520 0]{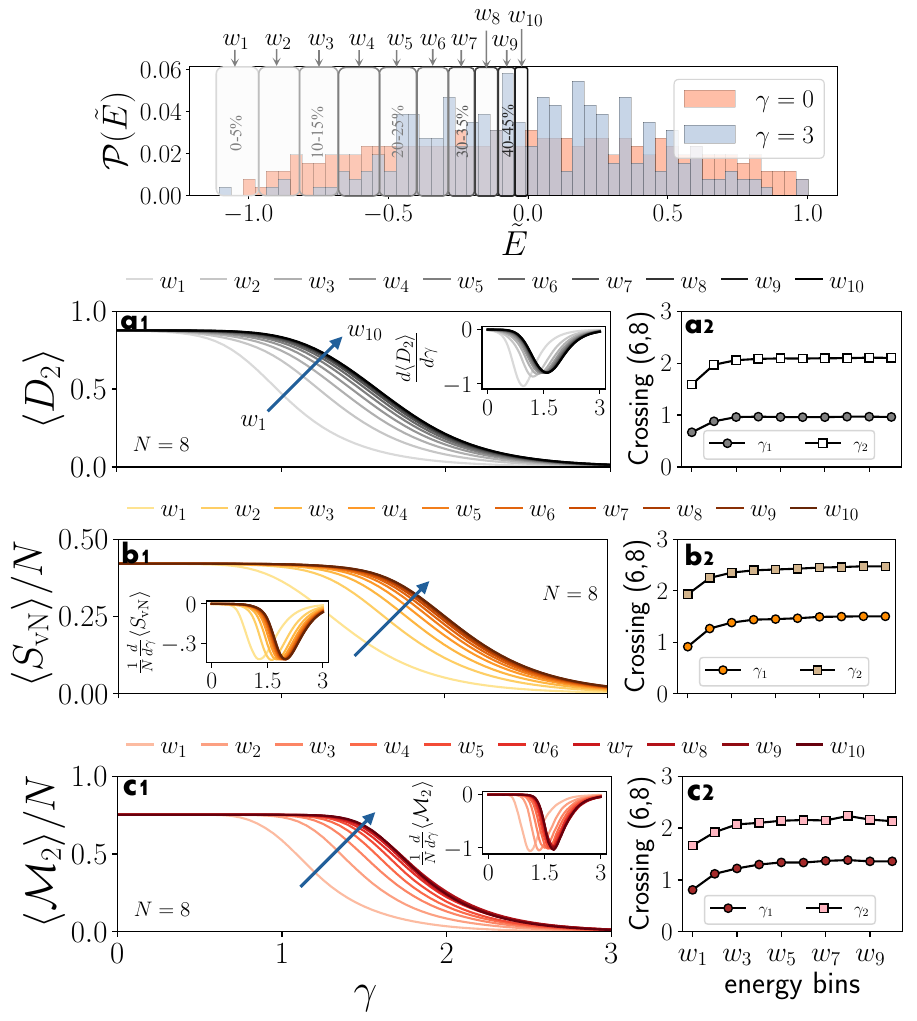}
    \caption{Dependence of the ergodic regime on energy window in the RP model. 
    Top panel: The eigenenergies span a broad range, with densely packed, nearly degenerate states located in the spectral bulk, while the edges are populated by relatively few states. In the ergodic phase ($\gamma=0$), the energy distribution is more flat as compared to the localized phase ($\gamma=3$). The eigenstates are ordered in ascending energy, and the lower half of the spectrum is divided into 10 equal windows, labeled $w_1$ to $w_{10}$, each containing 5\% of all eigenstates. 
    As the energy window shifts toward the spectral center (thin to thick lines), the $\mathcal{O}(1)$ ergodic plateau and corresponding transition points $\gamma_1$ and $\gamma_2$ move to higher $\gamma$ values for all complexity markers \textbf{(a1–c1)}. This shift saturates beyond approximately $w_3$ \textbf{(a2–c2)}, indicating that the crossing points remain essentially unchanged when averaging over bulk regions ranging from as narrow as the central 10\% to as broad as 90\% of the spectrum.
    Data shown for $N=6$ ($10000$ samples) and $N=8$ ($4000$ samples). Insets of \textbf{(a1-c1} show the derivative of the complexity markers, from which the crossing points are calculated.}

    \label{fig:edge_mobility}
\end{figure}

In Fig.~\ref{fig:edge_mobility}{\bf a1–c1}, we analyze consecutive $5\%$ segments of the energy spectrum. Since the spectrum is symmetric on average, we focus only on the lower half.
As we move toward the center of the spectrum, the transition curves systematically shift to the right, indicating that the transition point occurs at higher values of the control parameter, consistent with the presence of a mobility edge. Additionally, the curves corresponding to higher-energy windows quickly collapse onto one another, suggesting rapid convergence. This behavior supports our choice of using the central $20\%$ of the spectrum for the bulk calculations in section \ref{sec: RP_results}, as this range already displays stable and converged behavior.

The rapid convergence becomes even clearer when examining the evolution of the crossing points of the derivatives of the complexity markers, which we used in the main text to pinpoint the transitions. As shown in Fig.~\ref{fig:edge_mobility}{\bf a2–c2}, these crossing points quickly stabilize to a constant value above the energy window $10\%–15\%$, confirming the robustness of our choice.

\section{Scaling of crossing points}\label{Appendix: scaling of crossing}
To find the transition points in the thermodynamic limit, we fit a second-order polynomial curve to the crossing point with respect to $1/\sqrt{N_i N_{i+1}}$, of the form
\begin{equation}
     \gamma= a + b \frac{1}{\sqrt{N_i N_{i+1}}} + c \frac{1}{N_i N_{i+1}} 
\end{equation}
By extrapolating to $N\to \infty$, we obtain the transition points $\gamma^\infty=a$. 

To obtain the correct points after scaling, we found it crucial to set the acceptable parameter range of $b$ and $c$ while fitting the curve. For obtaining $\gamma_1$, the fitted curve should move to higher $\gamma$ with increasing $N$, i.e., $b,c \leq 0$, while for $\gamma_2$, it should move to smaller $\gamma$, i.e., $b,c \geq 0$. 

\section{Details on number of samples}\label{Appendix: number of samples}

For numerical reproducibility, we provide detailed tables listing the exact number of disorder realizations used for the results presented in the main text (see Tables~\ref{tab:fractal_dimension_samples}-\ref{tab:SRE_samples}). 
To keep the numerical effort manageable under the exponential growth of the Hilbert-space dimension with system size $N$, we progressively reduced the number of realizations with increasing $N$. 
However, this reduction does not compromise the accuracy of our results, as the observables become increasingly self-averaging with larger $N$ (see Fig.~\ref{fig:self_averaging}\textbf{a1,b1,c1}). 

Moreover, we adapted the number of samples for the marker under consideration. 
The computational complexity for computing the maximum entanglement grows as ${N \choose N/2} \times \text{construction}(\rho_A) \times \text{diag}(\rho_A) = \frac{2^{5/2N}}{\sqrt{N}}$. 
Similarly, the complexity for computing the Stabilizer Rényi Entropy (SRE) scales as $\mathcal{O}(2^{2N})$, due to the size of the Pauli group. 
In comparison, the computational complexity for calculating the fractal dimension is more modest, scaling as $\mathcal{O}(2^N)$, permitting us to increase the number of samples to even further improve the quality of the results. 
   
Among the considered models, we found the strongest sample-to-sample fluctuations in the PLRBM model, making a larger number of realizations necessary to achieve satisfactory statistical accuracy. 
In contrast, due to the large density of states in the bulk, a comparatively smaller number of realizations was sufficient for the RP bulk model as compared to studies on its ground state. 
For the SYK+Ising model, we maintained a comparable sample size to that of the RP bulk case across all markers.

\begin{table}[hbt!]
    \centering
    \begin{tabular}{l|rrrrrrrrr}
        \toprule
        \textbf{Model / N} & \textbf{4} & \textbf{5} & \textbf{6} & \textbf{7} & \textbf{8} & \textbf{9} & \textbf{10} & \textbf{11} & \textbf{12} \\ \hline 
        \midrule
        RP bulk     & 20000 & 15000 & 10000 & 5000  & 4000 & 3000 & 2000 & 1000 & 500 \\
        RP gs       & 20000 & 15000 & 10000 & 5000  & 4000 & 3000 & 2000 & 1000 & 500 \\
        PLRBM bulk  & 20000    & 15000    & 10000    & 5000    & 4000   & 3000   & 2000   & 1000   & 500 \\
        SYK+Ising   & 5000  & 4000  & 3000  & 2000  & 1000 & 500  & 100  & --   & -- \\
        \bottomrule
    \end{tabular}
    \caption{Number of samples used for fractal dimension estimates across different models and system sizes \(N\).}
    \label{tab:fractal_dimension_samples}
\end{table}

\begin{table}[hbt!]
    \centering
    \begin{tabular}{l|rrrrrrrrr}
        \toprule
        \textbf{Model / N} & \textbf{4} & \textbf{5} & \textbf{6} & \textbf{7} & \textbf{8} & \textbf{9} & \textbf{10} & \textbf{11} & \textbf{12} \\ \hline 
        \midrule
        RP bulk     & 5000  & 4000  & 3000 & 2000 & 1000 & 800  & 100 & 30 & -- \\
        RP gs       & 20000 & 15000 & 10000 & 5000 & 4000 & 3000 & 500 & 100 & 50 \\
        PLRBM bulk  & 35000    & 24000    & 40000   & 16000   & 10000   & 3000   & 500 & --  & -- \\
        SYK+Ising   & 5000  & 4000  & 3000 & 2000 & 1000 & 500  & 100 & --  & -- \\
        \bottomrule
    \end{tabular}
    \caption{Number of samples used for maximum entanglement entropy estimates across different models and system sizes \(N\).}
    \label{tab:entanglement_samples}
\end{table}

\begin{table}[hbt!]
    \centering
    \begin{tabular}{l|rrrrrrrrr}
        \toprule
        \textbf{Model / N} & \textbf{4} & \textbf{5} & \textbf{6} & \textbf{7} & \textbf{8} & \textbf{9} & \textbf{10} & \textbf{11} & \textbf{12} \\ \hline
        \midrule
        RP bulk     & 5000  & 4000  & 3000 & 2000 & 1000 & 500  & 80  & --  & -- \\
        RP gs       & 20000 & 15000 & 10000 & 5000 & 4000 & 1000 & 200 & 100 & 50 \\
        PLRBM bulk  & 35000    & 24000    & 20000   & 20000   & 10000   & 1500   & --  & --  & -- \\
        SYK+Ising   & 5000  & 4000  & 3000 & 2000 & 1000 & 500  & 100 & --  & -- \\
        \bottomrule
    \end{tabular}
    \caption{Number of samples used for Stabilizer Rényi Entropy estimates across different models and system sizes \(N\).}
    \label{tab:SRE_samples}
\end{table}

\section{Self-averaging property}
\label{Appendix: self-averaging}

In disordered quantum many-body systems, the self-averaging property refers to the tendency of observables to converge to their ensemble-averaged values as the system size increases. A physical quantity $\mathcal{C}$ is said to be self-averaging if its relative variance vanishes in the thermodynamic limit. This can be quantified by the dimensionless ratio
\begin{equation}
    R_{\mathcal{C}} = \frac{\mathrm{Var}(\mathcal{C})}{\langle \mathcal{C} \rangle^2} \,,
\end{equation}
where $\langle \cdot \rangle$ denotes the disorder average. In ergodic phases, where the system uniformly explores its Hilbert space, $R_{\mathcal{C}} \to 0$ as the system size increases, indicating strong self-averaging. In contrast, in non-ergodic or localized regimes, $R_{\mathcal{C}}$ can remain finite even in the thermodynamic limit, reflecting strong sample-to-sample fluctuations and the breakdown of ergodicity. 

\begin{figure*}[hbt!]
    \centering
    \includegraphics[width=\linewidth, clip, trim=0 80 230 25]{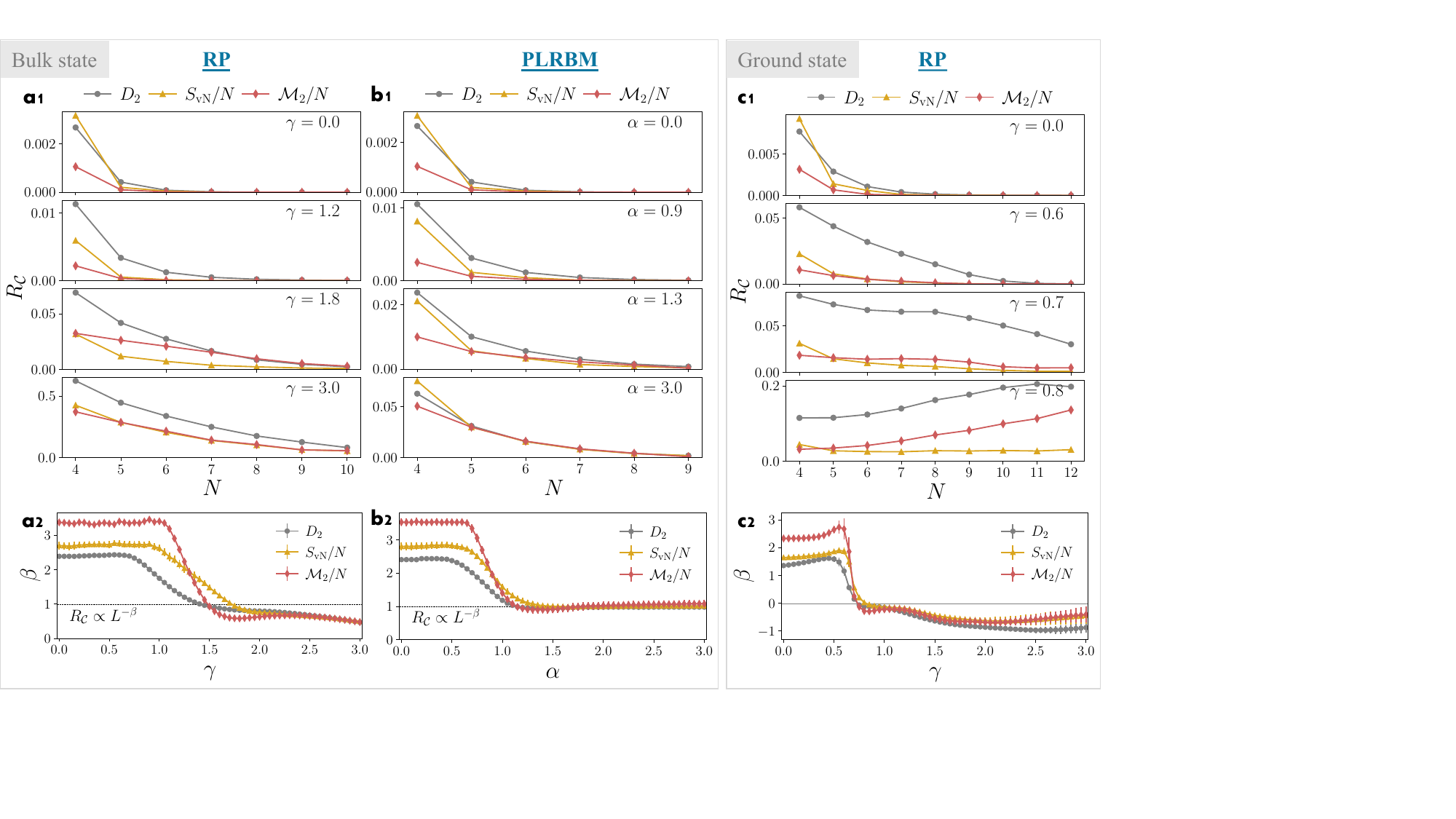}
    \caption{Self-averaging behavior of different complexity markers and their scaling exponents.
    \textbf{(a1)} For the bulk of the RP model, in the ergodic regime ($\gamma = 0$), all markers exhibit strong self-averaging, indicated by a vanishing relative variance. At intermediate disorder ($\gamma = 1.2$), $D_2$ already begins to deviate, showing non-self-averaging behavior, while others remain stable. At $\gamma = 1.8$ and $\gamma = 3$, none of the markers are self-averaging, consistent with the breakdown of ergodicity and the emergence of strong sample-to-sample fluctuations.
    \textbf{(a2)} Scaling exponent \(\beta\) as a function of \(\gamma\) for RP bulk states. All three markers exhibit strong self-averaging (\(\beta > 1\)) up to \(\gamma = 1.5\), and after \(\gamma = 2.5\), they show same $\beta$. Notably, \(\mathcal{M}_2\) displays the steepest variation in \(\beta\), indicating a more pronounced change in self-averaging across the transition. The deviation from the plateau region of constant \(\beta\) occurs in the same order, i.e., $D_2$ then $S_{\rm vN}$ and $\mathcal{M}_2$ as observed in Fig.~\ref{fig:RP_bulk_all}. 
    \textbf{(b1)} In the bulk states of the PLRBM model, complexity markers exhibit behavior analogous to the RP model, with magic showing the highest degree of self-averaging, followed by entanglement entropy, and fractal dimension displaying the least self-averaging in the ergodic phase. 
    \textbf{(b2)} The scaling exponent of all the markers becomes weakly self-averaging ($\beta\sim1$) close to $\alpha \sim 1$, and the $\alpha>1$ regime is comparatively more self-averaging than the bulk states of the RP model.  
    \textbf{(c1)} In the ground state of the RP model, all the markers stop exhibiting the self-averaging behavior after crossing the phase transition point $\gamma^\infty \sim 0.75$, leading to $\beta<0$ as seen in \textbf{(c2)}.}
    \label{fig:self_averaging}
\end{figure*}

\paragraph*{RP bulk states.}
In Fig.~\ref{fig:self_averaging}\textbf{a1}, we find that the complexity markers exhibit self-averaging behavior exclusively within the ergodic regime ($\gamma \leq 1$). Beyond this point, deviations begin to emerge. Notably, at $\gamma = 1.2$, the fractal dimension $D_2$ starts losing its self-averaging character due to its scaling as $\mathcal{O}(N^\delta)$, whereas both the normalized von Neumann entropy $S_{\rm vN}/N$ and the magic $\mathcal{M}_2/N$, which remain $\mathcal{O}(1)$, continue to exhibit self-averaging. At stronger disorder strengths, such as $\gamma = 1.8$, the entanglement remains more strongly self-averaging, consistent with a larger plateau in Fig.~\ref{fig:RP_bulk_all}\textbf{b1} of the $\mathcal{O}(1)$ regime. At $\gamma = 3$, none of the markers remain self-averaging, in accordance with their $\mathcal{O}(N^\delta)$ and $\mathcal{O}(1/N)$ scaling behaviors, respectively.

Furthermore, the strength of self-averaging can be inferred from the scaling behavior of \( R_\mathcal{C} \) with system size \( L = 2^N \), i.e.,
\begin{equation}
    R_\mathbf{C} \propto L^{-\beta} = 2^{-N\beta} \,.
\end{equation}
Systems are termed strongly self-averaging if \( \beta \geq 1 \), and weakly self-averaging when \( 0 < \beta < 1 \). Notably, Fig.~\ref{fig:self_averaging}\textbf{a2} shows that up to \( \gamma = 1.5 \), already in the non-ergodic phase, the system remains strongly self-averaging. Beyond \( \gamma = 2.5 \), all markers exhibit the same exponents. Additionally, \( \mathcal{M} \) demonstrates stronger self-averaging in the ergodic regime and undergoes the most pronounced change around the transition points.

\paragraph*{RP ground states.}
A key distinction between the bulk and ground state phases is the absence of a fractal regime in the latter, where only ergodic and localized phases are detected via complexity markers. Notably, the ground state exhibits a first-order phase transition, raising the question of whether self-averaging behaves differently compared to the bulk. As shown in Fig.~\ref{fig:self_averaging}\textbf{c1}, self-averaging holds at $\gamma = 0$, but breaks down beyond the transition point—at $\gamma = 0.8$, the marker even increases with $N$. This behavior is reminiscent of the vanishing marker values at large $\gamma$ (see Fig.~\ref{fig: gs_RP_plots}).

\paragraph*{PLRBM bulk states.}
Similar to the RP bulk states, the PLRBM bulk state also shows the strongly self-averaging behavior, and near the Anderson critical point, at $\alpha=1$, all the complexity markers still remain strongly self-averaging, i.e., $\beta=1$ (see Fig.~\ref{fig:self_averaging}\textbf{b2}), which is a behavior quite different from that in the RP bulk states.

Since the scaling behavior of the self-averaging property in complexity markers reflects the ergodicity transitions observed in the markers themselves, one might speculate that in disordered systems where fluctuations play a crucial role, the self-averaging property could serve as a diagnostic tool for phase transitions.

\section{SRE in Haar-random states}\label{Appendix: Pauli_&_Haar_random_states}
\begin{figure}[hbt!]
    \centering
    \includegraphics[width=0.65\columnwidth]{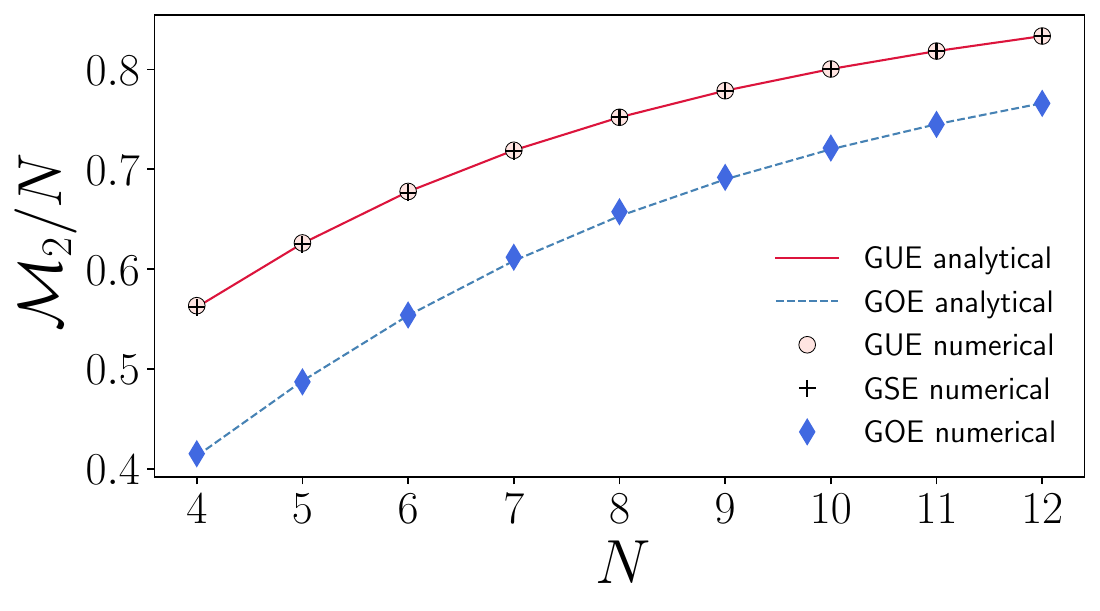}
    \caption{The magic computed for GUE and GSE ensembles exhibits similar behavior, while the GOE shows a distinct trend. The numerically obtained values are in excellent agreement with the corresponding analytical predictions.}
    \label{fig: analytical_gaussian_magic}
\end{figure}

For unitary-invariant Haar-random states, the expectation value of Pauli strings can be easily calculated analytically by averaging over Haar-random states.
Using the formula~\cite{Roberts:2016hpo}
\begin{equation}
    \int \ket{\psi}\bra{\psi}^{\otimes k} d \psi = \frac{(L-1)!}{(L+k-1)} \sum_{\pi \in S_k} W_\pi
\end{equation}
where $S_k$ is the permutation group and $W_\pi$ is the replica-permutation operator and $L=2^N$, we find
\begin{equation}
    \int |\bra{\psi} P_i \ket{\psi} |^2 d \psi = \begin{cases}
        1 & i = 0 ,\\
        \frac{1}{2^N+1} & i \neq 0 .
    \end{cases}
\end{equation}
The stabilizer Rényi entropy (SRE) can be calculated using similar techniques~\cite{turkeshi2025pauli} and takes the following analytical forms depending on the ensemble with respect to which the Haar measure is invariant:
\begin{equation}
    \begin{split}
        \mathcal{M}_2^{\rm Haar}(\text{GUE})=&\mathcal{M}_2^{\rm Haar}(\text{GSE})=-\log_2\left(\frac{4}{3+2^N}\right) \,  , \\
        \mathcal{M}_2^{\rm Haar}(\text{GOE})=&-\log_2\left(\frac{7}{6+2^N}\right) \,  .
    \end{split}
\label{Eq: Magic_RMT_analytical}
\end{equation}
These expressions match precisely with the numerically extracted $\mathcal{M}_2$ at $\gamma=0$ in the RP model and $\alpha=0$ in the PLRBM model, as illustrated in Fig.~\ref{fig: analytical_gaussian_magic}. In contrast, the pure SYK$_4$ model ($\lambda=0$) exhibits a deviation from the Haar-random SRE value.

\section{Pauli spectrum for RP and PLRBM bulk states}\label{Appendix: Pauli_spectrum}
The stabilizer Rényi entropy is defined as the Rényi entropy of a probability distribution constructed from the expectation values of Pauli strings on a given quantum state. Since a probability distribution inherently contains more information than any of its finite-order entropies, additional structure---particularly relevant in complex phases---is revealed by examining higher-order moments $ \mathcal{M}_q$.
Moreover, non-ergodic behavior is often captured by higher-order entropies, such as the fractal dimension $D_q$. 

\begin{figure}[hbt!]
    \centering
    \includegraphics[width=0.65\columnwidth, clip, trim= 5 170 400 0]{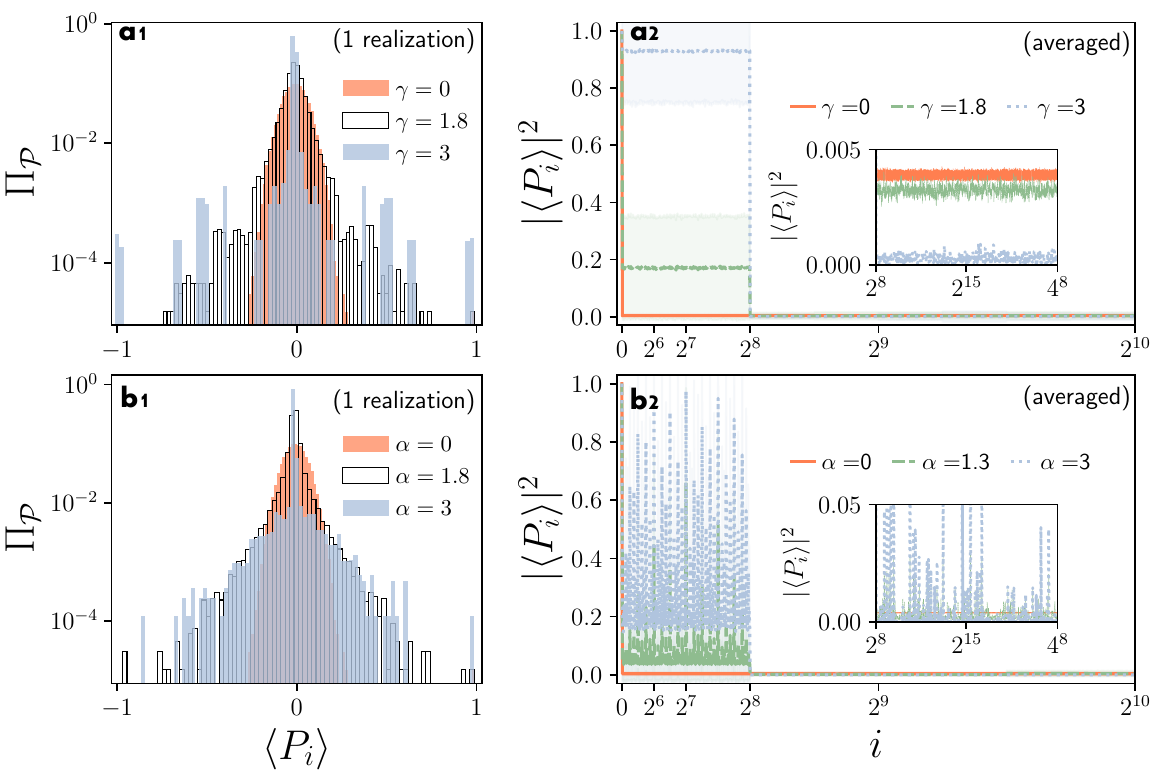}
    \caption{Frequency distribution for a single realization (left) and averaged Pauli-spectrum for a bulk state in the middle of the spectrum (right) of the RP (top) and PLRBM models (bottom). Data for $N=8$ with $4000$ and $8000$ realizations, respectively. In the ergodic phase (orange), expectation values are nearly uniformly distributed over all Pauli strings with minimal variance (shaded regions in \textbf{a2, b2}), resulting in a distribution with small variance around $\langle P_i \rangle =0$ in $\Pi_{\mathcal{P}}$ \textbf{(a1, b1)}. In contrast, localized states (blue) exhibit pronounced peaks in $\Pi_{\mathcal{P}}$, reflecting non-zero expectation values aligned with the stabilizer basis (notably within $i = [0, 2^N]$ in \textbf{a2, b2}). The intermediate regime shows partial delocalization with mixed features. These distinct distribution patterns are reflected in the magic values. Compared to RP, localization in PLRBM bulk states is weaker \textbf{(b1)} and exhibits reduced self-averaging behavior \textbf{(b2)}.}
    \label{fig:pauli_spectrum}
\end{figure}

It is therefore worth investigating if relevant signatures can emerge from the \emph{Pauli spectrum}, defined as the set of squared Pauli expectation values on the state~\cite{Beverland_2020, turkeshi2025pauli}:
\begin{equation}\label{eq: pauli_spectrum}
    {\rm spec}(\ket{\psi}) =\Big\{ |\langle \psi|P_i  |\psi\rangle|^2, \ P_i \in \mathcal{P}_N; \  i = 0,1, \dots ,4^N-1 \Big\} \,.
\end{equation}
The support of the Pauli spectrum, $i \in [0,4^N-1]$, changes depending on $N$. The \emph{frequency distribution} of these expectation values,
\begin{equation}\label{eq: frequency distribution}
    \Pi_{\mathcal{P}}(\ket{\psi})= \sum_{i} \frac{\delta(x-x_{P_i})}{4^N} \,, \qquad x_{P_i}=\langle \psi|P_i  |\psi\rangle ,
\end{equation}
offers a diagnostic tool for understanding the ergodicity behavior independent of $N$.
However, individual realizations would have different $\Pi_{\mathcal{P}}$, which can average away relevant features. 
Instead, we focus on the averaged Pauli spectrum, where the expectation value for each \( P_i \) is averaged over states from each realization. To better interpret its structure, we organize the Pauli strings by indexing $i$ in a specific order: the first $2^N$ indices \( i \in [0, 2^N - 1] \) correspond to diagonal operators \( \{ I, Z \}^{\otimes N} \), followed by all other (non-diagonal) strings.  The overall ordering is then refined by sorting the values of \( \langle P_i \rangle \) in descending order, using the localized state at \( \alpha = 10 \) or \( \gamma = 10 \) as a reference. This choice is motivated by the fact that localized states exhibit significant weight on the diagonal strings (see Fig.~\ref{fig:pauli_spectrum} {\bf a2}, {\bf b2}), and thus provide a natural order.

Ergodic states follow the Haar-random prediction (see Appendix~\ref{Appendix: Pauli_&_Haar_random_states}) and exhibit an almost uniform distribution over Pauli strings around the value $\frac{1}{2^N+1}$, except for the identity component $I^{\otimes N}$, which saturates to $1$, resulting in $\Pi_{\mathcal{P}}$ sharply peaked around zero (orange bars in Fig.~\ref{fig:pauli_spectrum} {\bf a1}, {\bf b1}) and low variance in the frequency distribution (orange shaded region in {\bf a2}, {\bf b2} ). In contrast, localized states have only a few significant Pauli components ({\bf a2}, {\bf b2}), leading to larger individual $\langle P_i \rangle$ values and a broadened spectrum (blue bars in {\bf a1}, {\bf b1}).

Notably, exact stabilizer states yield Pauli expectation values of either \( +1 \) for all \( 2^N \) stabilizers, or a balanced set of \( +1 \) and \( -1 \) values~\cite{turkeshi2025pauli}, as can be observed in the RP bulk case (blue lines in {\bf a2}). In the PLRBM model, however, bulk localized states do not align with the computational basis $\{I, Z\}^{\otimes N}$, resulting in a more dispersed Pauli spectrum ({\bf b1}) compared to the RP model ({\bf a1}). Even after averaging over random realizations, the PLRBM bulk state retains high variance in its Pauli spectrum across the non-ergodic and localized regimes, except in the ergodic phase where the distribution sharply concentrates~\cite{fabila2024reducing}.

In summary, the Pauli spectrum offers insight into why an ergodic state, which contains high magic, requires a larger number of basis states and is thus considered more complex, while a localized state, with non-zero contributing terms limited to a few Pauli strings, is considered less complex (see Fig.~\ref{fig: first_figure}\textbf{b}). Furthermore, the difference between the RP and PLRBM models is clearly captured in the Pauli spectrum, where the less suppressed off-diagonal terms of the PLRBM model yield strong fluctuations even in the localized regime, thus providing a direct interpretation of the relatively high magic in the localized regime of PLRBM bulk states (see Fig.~\ref{fig: RBM_results_bulk}\textbf{c1}), compared to its RP counterpart.

\section{Complexity markers for superposition of two computational basis states}\label{Appendix: SYK_magic}

It is instructive to examine the behavior of the three complexity markers for a simple superposition of two computational basis states. Consider the following $N$-qubit state:
\begin{equation}
    \ket{\psi}= \sin{\theta} \ket{a} + e^{i\phi} \cos{\theta} \ket{\bar{a}}\,,
\end{equation}
where $\ket{a}$ and $\ket{\bar{a}}$ are $N$-qubit orthogonal computational basis states differing by a Hamming distance of $N$, i.e., they are locally orthogonal at each qubit site. For such a state, the complexity markers are given by:
\begin{align}
    D_2 & = -\frac{1}{N} \log_2{\left(\sin^4{\theta} + \cos^4{\theta}\right)} \,, \\
    S_{\rm vN} & = -\left[ \cos^2{\theta} \log_2{\left(\cos^2{\theta}\right)} + \sin^2{\theta} \log_2{\left(\sin^2{\theta}\right)} \right] \,.
\end{align}
The SRE is given by~\cite{capecci2025}:
\begin{equation}
    \mathcal{M}_2 = -\log_2\left( \frac{1 + \cos^4{2\theta} + \sin^4{2\theta} \left( \cos^4{\phi} + \sin^4{\phi} \right)}{2} \right)\,.
\end{equation}

Among the three markers, only the SRE is sensitive to the relative phase $\phi$. In the special case of an equal superposition with $\pm 1$ difference, i.e., $\sin{\theta} = \cos{\theta} = \frac{1}{\sqrt{2}}$ and $\phi = 0, {\rm or \ } \pi $, the complexity markers simplify to
\begin{align}
    \label{eq:twobasisstatemarkerssimplified}
    D_2 = \frac{1}{N}, \quad \frac{S_{\rm vN}}{N} = \frac{1}{N}, \quad \frac{\mathcal{M}_2}{N} = 0\,.
\end{align}

Such a superposition state appears in the $\lambda\to1$ limit of the SYK+Ising model. Due to the fermion parity symmetry of the full Hamiltonian, the ground state has double degeneracy and forms a superposition of two degenerate states connected by $(-1)^F$:
\begin{equation}
    (-1)^F \ket{\rm GS_1} =+\ket{\rm GS_2}\,.
\end{equation}
Moreover, the symmetry ensures that they form equal superpositions without any complex relative phase, as explained in the following way. Consider the generic superposition
\begin{equation}
    \ket{\psi}=\sin{\theta} \ket{\rm GS_1} + e^{i\phi}\cos{\theta}\ket{\rm GS_2}\,.
\end{equation}
As $\ket{\psi}$ should also be symmetric under $(-1)^F$ with possible eigenvalues $\pm 1$,
\begin{equation}
    (-1)^F \ket{\psi}\stackrel{!}{=} \pm \ket{\psi} \Rightarrow \theta=\pi/4, \phi=0, \pi.
\end{equation}
Now, close to $\lambda \to 1$, as both ground states are computational basis states, we can use the above Eq.~\eqref{eq:twobasisstatemarkerssimplified}. 
That is, we expect fractal dimension and von-Neumann entropy density to scale as $1/N$ and SRE to vanish, which is exactly what we observe in Fig.~\ref{fig:SYK_results}\textbf{c}.

\section{Comparison between IPR and $D_2$}\label{appendix: ipr vs d2}

\begin{figure}[hbt!]
    \centering
    \includegraphics[width=0.65\columnwidth]{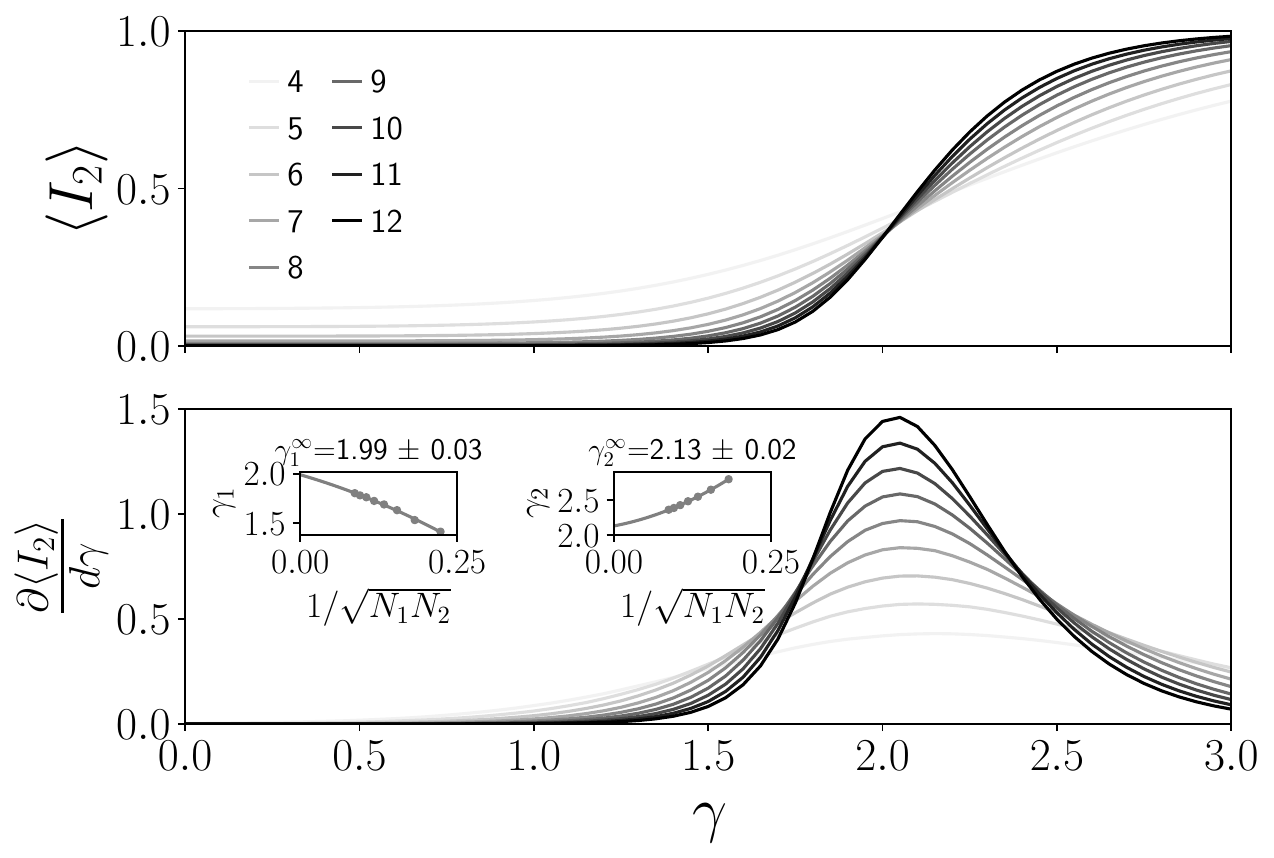}
    \caption{Unlike the fractal dimension, the average IPR does not resolve two distinct transition points in the bulk phase diagram of the RP model (Fig.~\ref{fig:RP_bulk_all}). Instead, the derivative crossings yield nearly coincident transition points, suggesting a single transition point at $\gamma_c \simeq 2$.}
    \label{fig:IPRvsD2}
\end{figure}

Although the fractal dimension $D_2$ can be formally related to the inverse participation ratio (IPR) $I_2$, defined in Eq.~\eqref{eq:IPR}, via the expression $D_q = \frac{\log_2(I_q)}{N(1 - q)}$, relying solely on $I_q$ to detect phase transitions in such models can be challenging. This is because the multifractal nature of the wavefunctions manifests through the scaling relation $I_q \propto 1/L^{D_q(q-1)}$~\cite{castellani1986multifractal, mirlin1996transition,Mirlin_2000, evers2000fluctuations,Cadez_2024}, which can be captured in $D_q$ instead of $I_q$. As illustrated in Fig.~\ref{fig:IPRvsD2}, the averaged IPR $\langle I_2 \rangle$ reveals only the transition near $\gamma \simeq 2$, corresponding to the non-ergodic to localized transition, while it fails to resolve the ergodic to non-ergodic transition. This limitation highlights the advantage of using the fractal dimension as a more sensitive diagnostic of intermediate, multifractal phases. Moreover, taking the logarithm of the IPR is known to generally reduce the sample-to-sample fluctuations and improve the self-averaging behavior~\cite{Santos_self_averaging}.

\end{appendix}

\end{document}